\documentclass[fleqn,usenatbib]{mnras}

\usepackage[T1]{fontenc}

\DeclareRobustCommand{\VAN}[3]{#2}
\let\VANthebibliography\thebibliography
\def\thebibliography{\DeclareRobustCommand{\VAN}[3]{##3}\VANthebibliography}

\usepackage{graphicx}	
\usepackage{latexsym,amssymb,amsbsy,amsmath,natbib}
\usepackage{nccmath}
\usepackage{newtxtext,newtxmath}
\usepackage{array}
\usepackage{threeparttable}

\title[The SCUBA-2 Large eXtragalactic Survey: XMM-LSS]{The SCUBA-2 Large eXtragalactic Survey: $\mathbf{850}\,\boldsymbol{\mu}{\rm {\mathbf m}}$ map, catalogue and the bright-end number counts of the XMM-LSS field}
\author[T. K. Garratt et al.]{
T. K. Garratt,$^{1}$\thanks{E-mail: t.garratt@herts.ac.uk}
J. E. Geach,$^{1}$
Y. Tamura,$^{2}$
K. E. K. Coppin,$^{1}$
M. Franco,$^{1}$
Y. Ao,$^{3,4}$
C. -C. Chen,$^{5}$
\newauthor
C. Cheng,$^{6}$
D. L. Clements,$^{7}$
Y. S. Dai,$^{6,8}$
H. Dannerbauer,$^{9,10}$
T. R. Greve,$^{11,12}$
B. Hatsukade,$^{13}$
\newauthor
H. S. Hwang,$^{14,15}$
L. Jiang,$^{16}$
K. Kohno,$^{13}$
M. P. Koprowski,$^{17}$
M. J. Micha\l{}owski,$^{18}$
M. Sawicki,$^{19}$
\newauthor
D. Scott,$^{20}$
H. Shim,$^{21}$
T. T. Takeuchi,$^{2,22}$
W. -H. Wang,$^{5}$
Y. Q. Xue,$^{23,4}$
C. Yang,$^{24}$
\\
$^{1}$Centre for Astrophysics Research, University of Hertfordshire, Hatfield, AL10 9AB, UK.\\
$^{2}$Division of Particle and Astrophysical Science, Nagoya University, Nagoya, Aichi 464-8602, Japan\\
$^{3}$Purple Mountain Observatory and Key Laboratory for Radio Astronomy, Chinese Academy of Sciences, 10 Yuanhua Road, Nanjing 210023, China \\
$^{4}$School of Astronomy and Space Science, University of Science and Technology of China, Hefei 230026, Anhui, China\\
$^{5}$Academia Sinica Institute of Astronomy and Astrophysics, No. 1, Section 4, Roosevelt Road, Taipei 10617, Taiwan\\
$^{6}$Chinese Academy of Sciences South America Center for Astronomy, National Astronomical Observatories, CAS, Beijing 100101, China\\
$^{7}$Imperial College London, Blackett Lab, Prince Consort Road, London, SW7 2AZ, UK\\
$^{8}$National Astronomical Observatories, Chinese Academy of Sciences, 20A Datun Road,Chaoyang District, Beijing 100101, China\\
$^{9}$Instituto de Astrof\'isica de Canarias (IAC), E-38205 La Laguna, Tenerife, Spain\\
$^{10}$Universidad de La Laguna, Dpto. Astrof\'isica, E-38206 La Laguna, Tenerife, Spain\\
$^{11}$National Space Institute, DTU Space, Technical University of Denmark, Elektrovej 327, DK-2800 Kgs. Lyngby, Denmark\\
$^{12}$Department of Physics and Astronomy, University College London, Gower Street, London WC1E 6BT, UK\\
$^{13}$Institute of Astronomy, Graduate School of Science, The University of Tokyo, 2-21-1 Osawa, Mitaka, Tokyo 181-0015, Japan\\
$^{14}$Astronomy Program, Department of Physics and Astronomy, Seoul National University , 1 Gwanak-ro, Gwanak-gu, Seoul 08826, Republic of Korea\\
$^{15}$ SNU Astronomy Research Center, Astronomy Program, Seoul National University, 1 Gwanak-ro, Gwanak-gu, Seoul 08826, Republic of Korea\\
$^{16}$Kavli Institute for Astronomy and Astrophysics, Peking University, No. 5 Yiheyuan Road, Haidian District,Beijing 100871, China\\
$^{17}$Institute of Astronomy, Faculty of Physics, Astronomy and Informatics, Nicolaus Copernicus University, Grudziadzka 5, 87-100 Torun, Poland\\
$^{18}$Astronomical Observatory Institute, Faculty of Physics, Adam Mickiewicz University, ul. S\l{}oneczna 36, 60-286 Pozna\'{n}, Poland\\
$^{19}$Department of Astronomy and Physics, and Institute for Computational Astrophysics, Saint Mary’s University, Halifax, NS B3H 3C3, Canada\\
$^{20}$Department of Physics and Astronomy, University of British Columbia, Canada\\
$^{21}$Department of Earth Science Education, Kyungpook National University, Daegu 41566, Republic of Korea\\
$^{22}$The Research Center for Statistical Machine Learning, the Institute of Statistical Mathematics, 10--3 Midori-cho, Tachikawa, Tokyo 190--8562, Japan\\
$^{23}$CAS Key Laboratory for Research in Galaxies and Cosmology, Department of Astronomy, University of Science and Technology of China, Hefei 230026, China\\
$^{24}$Department of Space, Earth and Environment Chalmers, University of Technology, G\"{o}teborg, Sweden\\
}
\date{Accepted XXX. Received YYY; in original form ZZZ}

\pubyear{2022}

\begin{document}
\label{firstpage}
\pagerange{\pageref{firstpage}--\pageref{lastpage}}
\maketitle
\begin{abstract} 
 
 \noindent We present $850\,\mu{\rm m}$ imaging of the XMM-LSS field observed for $170$ hours as part of the James Clerk Maxwell Telescope SCUBA-2 Large eXtragalactic Survey (S2LXS). S2LXS XMM-LSS maps an area of $9\,{\rm deg}^2$, reaching a moderate depth of $1\sigma \simeq 4\,{\rm mJy\,beam^{-1}}$. This is the largest contiguous area of extragalactic sky mapped by JCMT at $850\,\mu{\rm m}$ to date. The wide area of the S2LXS XMM-LSS survey allows us to probe the ultra-bright ($S_{\rm 850\mu m}\gtrsim15\,{\rm mJy}$), yet rare submillimetre population. We present the S2LXS XMM-LSS catalogue, which comprises $40$ sources detected at ${>}5\sigma$ significance, with deboosted flux densities in the range of $7\,{\rm mJy}$ to $48\,{\rm mJy}$. We robustly measure the bright-end of the $850\,\mu{\rm m}$ number counts at flux densities ${>}7\,{\rm mJy}$, reducing the Poisson errors compared to existing measurements. The S2LXS XMM-LSS observed number counts show the characteristic upturn at bright fluxes, expected to be motivated by local sources of submillimetre emission and high-redshift strongly lensed galaxies. We find that the observed $850\,\mu{\rm m}$ number counts are best reproduced by model predictions that include either strong lensing or source blending from a 15 arcsec beam, indicating that both may make an important contribution to the observed over-abundance of bright single dish $850\,\mu{\rm m}$ selected sources. We make the S2LXS XMM-LSS $850\,\mu{\rm m}$ map and ${>}5\sigma$ catalogue presented here publicly available. 
\end{abstract}




\section{Introduction}

Historically our knowledge of how galaxies form and evolve has been driven by observations made at optical wavelengths.  However, over the past $50$ years, helped by the boom in infrared/millimetre facilities (both ground-based and in space), the study of the Universe at infrared/millimetre wavelengths has quickly gained momentum.  The first infrared observations targeted a small number of known extragalactic sources revealing that some galaxies emit the same amount of energy in the infrared as at optical wavelengths \citep[e.g.,][]{Low1968,Kleinmann1970}.  Over the course of the following decade the scale of infrared surveys dramatically expanded, with the first all-sky infrared survey \citep[the \textit{Infrared Astronomical Satellite} mission;][]{Neugebauer1984} detecting ${\sim}20{,}000$ local starburst galaxies, the majority of which were too faint to be included in previous optical catalogues \citep{beichmann1988}. In the 1990's the NASA \textit{Cosmic Background Explorer} \citep{boggess1992} gave us the first measurements of the Cosmic Infrared Background (CIB; the integrated infrared emission from all galaxies in the history of the Universe).  These data \citep[e.g,][]{Puget1996,Fixsen1998,Hauser1998}, along with measurements of the CIB at mid-infrared wavelengths \citep[e.g.,][]{Papovich2004,Dole2006} revealed that the Extragalactic Background Light has comparable intensities at optical ($\lambda<8\mu{\rm m}$) and infrared ($\lambda>8\mu{\rm m}$) wavelengths, meaning that around half the optical and ultra-violet emission from galaxies is absorbed by interstellar dust and re-emitted in the far-infrared \citep[e.g., ][]{Dole2006,Hauser2001}.  

The most luminous infrared galaxies are known as Ultra Luminous Infrared Galaxies (ULIRGs) and Hyper Luminous Infrared Galaxies (HyLIRGs) with integrated infrared luminosities of $L_{\rm IR}>10^{12}\,{\rm L}_{\odot}$ and  $L_{\rm IR}>10^{13}\,{\rm L}_{\odot}$ respectively, and star formation rates commmonly in excess of $50\,{\rm M}_{\odot}{\rm yr}^{-1}$ \citep[e.g.,][]{Sanders1996,casey2014}.  In these dusty star-forming galaxies light from young, short-lived, massive stars is absorbed by the surrounding dust and re-radiated at infrared wavelengths. 

Dusty star forming galaxies detected at submillimetre wavelengths are commonly known as Submillimetre Galaxies (SMGs). 
Since the first observations of SMGs in the late 1990's \citep[e.g.,][]{smail1997,barger1998,Hughes1998} there has been considerable progress in our understanding of their physical properties and cosmological significance in the context of galaxy evolution \citep[see][for detailed reviews]{casey2014,Hodge2020}.  We now know that SMGs selected at $850\,\mu{\rm m}$ are intrinsically highly infrared luminous ($L_{\rm IR}>10^{12}\,{\rm L}_{\odot}$) with star formation rates in excess of $100\,{\rm M}_{\odot}{\rm yr}^{-1}$ \citep[e.g.,][]{Chapman2005,magnelli2012,swinbank2014,ikarashi2015,michalowski2017,Miettinen2017}, have a number density which peaks at $\langle z\rangle \approx 2$--$3$ \citep[e.g.,][]{Chapman2005,Pope2005,wardlow2011,simpson2014,Chen2016,Miettinen2017}, have high stellar masses \citep[$M_{*}>10^{10}\,{\rm M}_\odot$; e.g.,][]{swinbank04,hainline2011,Michalowski2012,da_cunha2015,michalowski2017}, are gas-rich \citep[e.g.,][]{Frayer1998,Greve2005,Tacconi2006,Ivison2011,thomson2012,bothwell2013}, often host Active Galactic Nuclei \citep[e.g.,][]{Alexander2005,Johnson2013,Pope2008,Wang2013}, and with a volume density three orders of magnitude greater than that of local ULIRGS \citep[e.g.,][]{smail1997,Chapman2005} contribute approximately $20\%$ of the total star formation rate density over a redshift range $1$ to $4$ \citep[e.g.,][]{Casey2013,swinbank2014}.  These properties make SMGs excellent candidates for the progenitor population of massive elliptical galaxies seen in the local Universe today.  

Submillimetre astronomy benefits from a strong negative K correction that arises from the shape of the intrinsic spectral energy distribution (SED) of a typical star-forming galaxy, which peaks at $100\,\mu{\rm m}$. When we observe emission from a local star-forming galaxy at a wavelength of $850\,\mu{\rm m}$ we are probing the Rayleigh-Jean's tail (the rest-frame cold dust continuum emission) of the galaxy's intrinsic SED, and so, as we observe star-forming galaxies at increasing cosmological distances the rest-frame emission we observe at $850\,\mu{\rm m}$ moves up the Rayleigh-Jeans tail, closer to the peak of the intrinsic SED. The increasing power of rest-frame emission with increasing cosmological distance compensates for cosmological dimming.  This means that submillimetre galaxies of equal luminosity that lie at $1<z<8$ will have similar observed fluxes at $850\,\mu{\rm m}$, making SMGs a useful probe of the early Universe.  

The first extragalactic submillimetre surveys were conducted with the Submillimeter Common User Bolometer Array camera \citep[SCUBA;][]{Holland1999} mounted on the James Clerk Maxwell telescope (JCMT), providing the first census of dust-obscured star-formation in distant galaxies at submillimetre wavelengths \citep{smail1997,Hughes1998,barger1998}. Subsequent extragalactic submillimetre surveys built on the success of these early observations, and aided by the development of new submillimetre/millimetre instruments  such as the Large Apex BOlometer Camera Array \citep[LABOCA;][]{Siringo2009}, AzTEC \citep{Wilson2008} and the MAx-Planck Millimeter BOlometer \citep[MAMBO][]{Kreysa1998}, the scale of extragalactic surveys at submillimetre/millimetre wavelengths expanded \citep[e.g.,][]{Coppin2006,Eales2000,Scott2008,Weiss2009}.  However, the limited field of view and sensitivity of the early submillimetre/millimetre instruments made it difficult to map large areas of the sky (i.e., $>1\,{\rm deg}^2$), with the largest and deepest of these surveys detecting around $100$ sources \citep{Coppin2006,Weiss2009}, insufficient numbers for robust statistical studies of this cosmologically important population of galaxies.

A breakthrough in single-dish submillimetre astronomy came with the introduction of SCUBA-2 \citep{Holland2013}, the second generation bolometer array on the JCMT.  SCUBA-2 is a dual wavelength camera with two arrays of $5120$ pixels that simultaneously map the sky at wavelengths of $450\,\mu{\rm m}$ and $850\,\mu{\rm m}$, covering a field of view of $8\,{\rm arcmin}^2$.  With a mapping speed that is over an order of magnitude faster than that of its predecessor (at equivalent depth), SCUBA-2 opened the door for wide area (i.e., $>1\,{\rm deg}^2$) surveys at submillimetre wavelengths. The first such survey was the SCUBA-2 Cosmology Legacy Survey \citep[S2CLS;][]{Geach2017}, which began shortly after the commissioning of SCUBA-2.  This survey, the largest of seven JCMT Legacy Survey programs, mapped a total area of $5\,{\rm deg}^2$ over seven extragalactic fields to a median depth (at $850\,\mu{\rm m}$) of $1\sigma\simeq1\,{\rm mJy}\,{\rm beam}^{-1}$.  Owing to the wide-area and depth of S2CLS, this survey detected almost $3000$ sources at a significance of ${>}3.5\sigma$ in the $850\,\mu{\rm m}$ maps, a sample an order of magnitude larger than in previous submillimetre surveys, allowing the single dish $850\,\mu{\rm m}$ number counts to be measured to unprecedented accuracy.  The S2CLS $850\,\mu{\rm m}$ number counts are well fit by a \citet{Schechter1976} function of the form

\begin{ceqn}
\begin{align}
\frac{dN}{dS} = \left(\frac{N_0}{S_0}\right)\left(\frac{{S}}{S_0}\right)^{-\gamma}{\rm exp}\left(-\frac{{S}}{S_0}\right), 
\label{eq: schechter}
\end{align}
\end{ceqn}

\noindent with $N_0=7180\pm1200\,{\rm deg}^{-2}$, $S_0=2.5\pm0.4\,{\rm mJy}\,{\rm beam}^{-1}$ and $\gamma = 1.5\pm0.4$.  However, at intrinsic fluxes above $15\,{\rm mJy}$ there is a clear upturn in the S2CLS source counts, which lie above the \citet{Schechter1976} function fit. An over-abundance of bright sources is also seen at far-infrared wavelengths in the \textit{Herschel} Astrophysical Terahertz Large Area Survey \citep[H-ATLAS;][]{Eales2010, negrello2010} and the \textit{Herschel} Multi-tiered Extragalactic Survey \citep[HerMES;][]{Oliver2012,wardlow2013}, and at millimetre wavelengths (but only at high flux densities, $S_{1.4\,{\rm mm}}\gtrsim10\,{\rm mJy}$, corresponding to $S_{850\mu{\rm m}}\gtrsim40\,{\rm mJy}$) in the South Pole Telescope Sunyaev Zel'dovich survey \citep[SPT-SZ;][]{Vieira2010,Mocanu2013}, and is attributed to the presence of local objects and high-redshift gravitationally lensed sources.  It has been demonstrated that a simple flux density cut at $100\,{\rm mJy}$ in the \textit{Herschel} $500\,\mu{\rm m}$ band is almost $100$ per cent effective at selecting strongly lensed galaxies, after local ($z<0.1$) sources of emission have been removed \citep[][]{negrello2010,wardlow2013}.  S2CLS \citep{Geach2017} provides tentative evidence that a flux cut at $850\,\mu {\rm m}$ may yield a similar result -- the brightest $850\,\mu{\rm m}$ selected sources in this survey are a lensed high-redshift galaxy \citep[`Orochi';][]{Ikarashi2011} and a well-known Galactic object (the Cat's Eye Nebula).
However, with only six $850\,\mu{\rm m}$ selected sources in S2CLS with intrinsic fluxes above $15\,{\rm mJy}$, the bright end of the number counts remains poorly constrained and the potential of a flux cut at this wavelength to select strongly lensed galaxies barely explored. The brightest submillimetre galaxies selected at $850\,\mu{\rm m}$ are also observed to be the most distant \citep[e.g.,][]{Stach2019,Simpson2020,Chen2022}, but with so few single dish selected $850\,\mu{\rm m}$ sources with flux densities $S_{850\,\mu{\rm m}}>15\,{\rm mJy}$ the redshift distribution of sources in this flux regime is poorly constrained, and the trend of increasing redshift with submillimetre flux density is untested for these ultra-bright sources.  Intrinsically bright submillimetre sources are rare, with abundances for sources with flux densities ${>}20\,{\rm mJy}$ currently estimated to be only ${\approx}0.5$--$5$ galaxies per square degree \citep[e.g.,][]{bethermin2012,Geach2017}, and so a moderate depth survey at $850\,\mu{\rm m}$ with an area ${>10}\,{\rm deg}^2$ is needed to properly probe this far-infrared luminous, and potentially high redshift population of galaxies. 

The SCUBA-2 Large eXtragalactic Survey (S2LXS) is a JCMT Large Program (PI: J. E. Geach and Y. Tamura) and covers an area of $10\,{\rm deg}^2$ split over two fields - the X-ray Multi-Mirror Large Scale Structure Survey field \citep[XMM-LSS;][]{Pierre2004} and the Extended Cosmic Evolution Survey field \citep[E-COSMOS;][]{Scoville2007}.  In the context of other JCMT Large Programs S2LXS can be seen as a wide, moderately deep tier, complementing the SCUBA-2 Cosmology Legacy Survey \citep[S2CLS;][]{Geach2017}, the SCUBA-2 COSMOS survey \citep[S2COSMOS;][]{Simpson2019} and the SCUBA-2 Ultra Deep Imaging East Asian Observatory Survey \citep[STUDIES;][]{Wang2017}, which map smaller areas to higher sensitivity. S2LXS covers an area $2\times$ larger than S2CLS, is effectively $2\times$ deeper than SPT-SZ \citep{Vieira2010, Mocanu2013} and at a wavelength of $850\,\mu{\rm m}$ is more sensitive to dust emission from high redshift ($z\gtrsim4$) SMGs than far-infrared surveys \citep[i.e., H-ATLAS, HerMES;][]{Eales2010,Oliver2012}.  The primary aims of S2LXS are to detect the most luminous and highest redshift submillimetre sources, thus providing robust constraints on the bright-end of the single-dish $850\,\mu{\rm m}$ number counts and unveiling a new distant population of SMGs.

In this paper we present the results of the S2LXS XMM-LSS field and provide the map and source catalogue for public use. This paper is organised as follows: in section 2 we define the survey and describe the data reduction; in section 3 we present the maps and describe the cataloguing procedure; in section 4 we use these data to measure the bright-end of the $850\,\mu{\rm m}$ number counts and compare to recent results from model predictions; and in section $5$ we summarize the paper.

\section{The SCUBA-2 Large eXtragalactic Survey }
The SCUBA-2 Large eXtragalactic Survey (S2LXS) is a wide area (${>}10\,{\rm deg}^2$) survey at $850\,\mu{\rm m}$ split over two fields: XMM-LSS ($9\,{\rm deg}^2$); and E-COSMOS ($3\,{\rm deg}^2$). These fields were selected to take advantage of the wealth of legacy multi-wavelength data available (this data is essential for identifying SMG counterparts) and for the accessibility to key submillimetre/millimetre interferometers for follow-up observations. The S2LXS observations of the E-COSMOS field are currently less than $50$ per cent complete and continue under programme ID M20AL026 (PI: J. E. Geach and Y. Tamura). The S2LXS map, catalogue and results for the E-COSMOS field will be the subject of a future paper. Therefore the focus of this paper is the S2LXS XMM-LSS field, which is mapped at 850\,$\mu{\rm m}$ to a median depth of $1\sigma \simeq 4\,{\rm mJy\,beam}^{-1}$. Although SCUBA-2 simultaneously collects data at both 850\,$\mu{\rm m}$ and 450\,$\mu{\rm m}$ we do not process the complementary 450\,$\mu{\rm m}$ data. At 450\,$\mu{\rm m}$ the estimated sensitivity is $1\sigma\simeq 50\,{\rm mJy\,beam}^{-1}$ and so these data are not expected to have sufficient depth to reliably detect (at ${>}3.5\sigma$) even the most luminous submillimetre sources in the S2LXS XMM-LSS field.


XMM-LSS benefits from being a well-studied deep field in the Hyper Suprime-Cam (HSC) Subaru Strategic Program (HSC-SSP). The HSC-SSP is multi-band (\textit{grizy}+narrow-band) imaging survey \citep{Aihara2018} conducted using the HSC digital imaging camera on the $8.2\,{\rm m}$ Subaru Telescope. This survey comprises three tiers; a Wide tier ($1400\,{\rm deg}^2$, $r_{\rm AB}\simeq26$), a Deep tier ($28\,{\rm deg}^2$, $r_{\rm AB}\simeq27$) and an Ultra-Deep tier ($13.5\,{\rm deg}^2$, $r_{\rm AB}\simeq28$). The XMM-LSS field is also covered by legacy datasets at X-ray, ultra-violet, infrared and radio wavelengths (see Table \ref{tab:multi-wavelength data} for a summary of the multi-wavelength imaging available). Multi-wavelength data is essential for identifying counterparts and companions (e.g., satellite galaxies that are not dusty) to sources detected at $850\,\mu{\rm m}$.

\begin{table*}
	\centering
	\caption{Summary of ancillary multi-wavelength data for the S2LXS XMM-LSS field. Note that a region of this field (central coordinates $\alpha = 34^{\circ}.4542$, $\delta = -5^{\circ}.0986$) was observed at $850\,\mu{\rm m}$ in S2CLS \citep{Geach2017}. Observations of this region are not repeated in S2LXS.} 
	\label{tab:multi-wavelength data}
	\begin{tabular}{>{\centering\arraybackslash}m{7.5cm}>{\centering\arraybackslash}m{2.0cm}>{\centering\arraybackslash}m{3.7cm}>{\centering\arraybackslash}m{2.8cm}} 
	\hline \\ [-4pt]
		 Survey/Instrument & Abbreviation & Band & Reference \\[4pt]
		\hline \\ [-4pt]
                X-ray Multi-Mirror Large Scale Structure survey & XMM-LSS & [$0.5$-$2$]\,keV & \citet{Pierre2004} \\[4pt]
		X-ray Multi-Mirror \textit{Spitzer} Extragalactic Representative Volume Survey & XMM-SERVS & [$0.5$-$10$]\,keV & \citet{Chen2018} \\[9pt]
Canada-France-Hawaii Telescope Large Area \textit{U}-band Deep Survey & CFHT CLAUDS & $u^*$ & \citet{Sawicki2019} \\[4pt] 
Hyper Suprime-Cam Subaru Strategic Program & HSC-SSP & \textit{grizy} & \citet{Aihara2018} \\[4pt] 
Visible and Infrared Survey Telescope for Astronomy \newline Deep Extragalactic Observations survey & VISTA VIDEO & \textit{YJHK} & \citet{Jarvis2013} \\[9pt]
United Kingdom infrared telescope Infrared Deep Sky Survey Deep eXtragalactic Survey & UKIDSS DXS & \textit{YJHK} & \citet{Lawrence2007} \\[9pt]
\textit{Spitzer} Extragalactic Representative Volume Survey & SERVS &$3.6\,\mu{\rm m}$, $4.5\,\mu{\rm m}$ & \citet{Mauduit2012} \\[4pt]
\textit{Spitzer} survey of the Deep Drilling Fields & Deep Drill &$3.6\,\mu{\rm m}$, $4.5\,\mu{\rm m}$ & \citet{lacy2021} \\[4pt]
\textit{Spitzer} Wide-area InfraRed Extragalactic survey & SWIRE & $3.6\,\mu{\rm m}$, $4.5\,\mu{\rm m}$, $5.8\,\mu{\rm m}$, $8.0\,\mu{\rm m}$, $24\,\mu{\rm m}$, $70\,\mu{\rm m}$, $160\,\mu{\rm m}$ & \citet{Lonsdale2003} \\[9pt]
\textit{Herschel} Multi-tiered Extragalactic Survey & HerMES & $250\,\mu{\rm m}$, $350\,\mu{\rm m}$, $500\,\mu{\rm m}$ & \citet{Oliver2012} \\[4pt]
LOw Frequency ARray & LOFAR & $120$-$168$\,{\rm MHz}& \citet{Hale2019} \\[4pt]
MeerKAT International GigaHertz Tiered Extragalactic Explorations & MIGHTEE & $856$-$1712\,{\rm MHz}$&\citet{Heywood2022,jarvis2016}\\[9pt]
Very Large Array $1$-$2\,{\rm GHz}$ survey & -- & $1$--$2\,{\rm GHz}$ & \citet{heywood2020}\\[4pt] 
SCUBA-2 Cosmology Legacy Survey & S2CLS & $450\,\mu{\rm m}$, $850\,\mu{\rm m}$ & \citet{Geach2017}\\[4pt] 
		\hline
	\end{tabular}
\end{table*}

\subsection{Observations} 
\begin{figure}
	\includegraphics[width=\columnwidth]{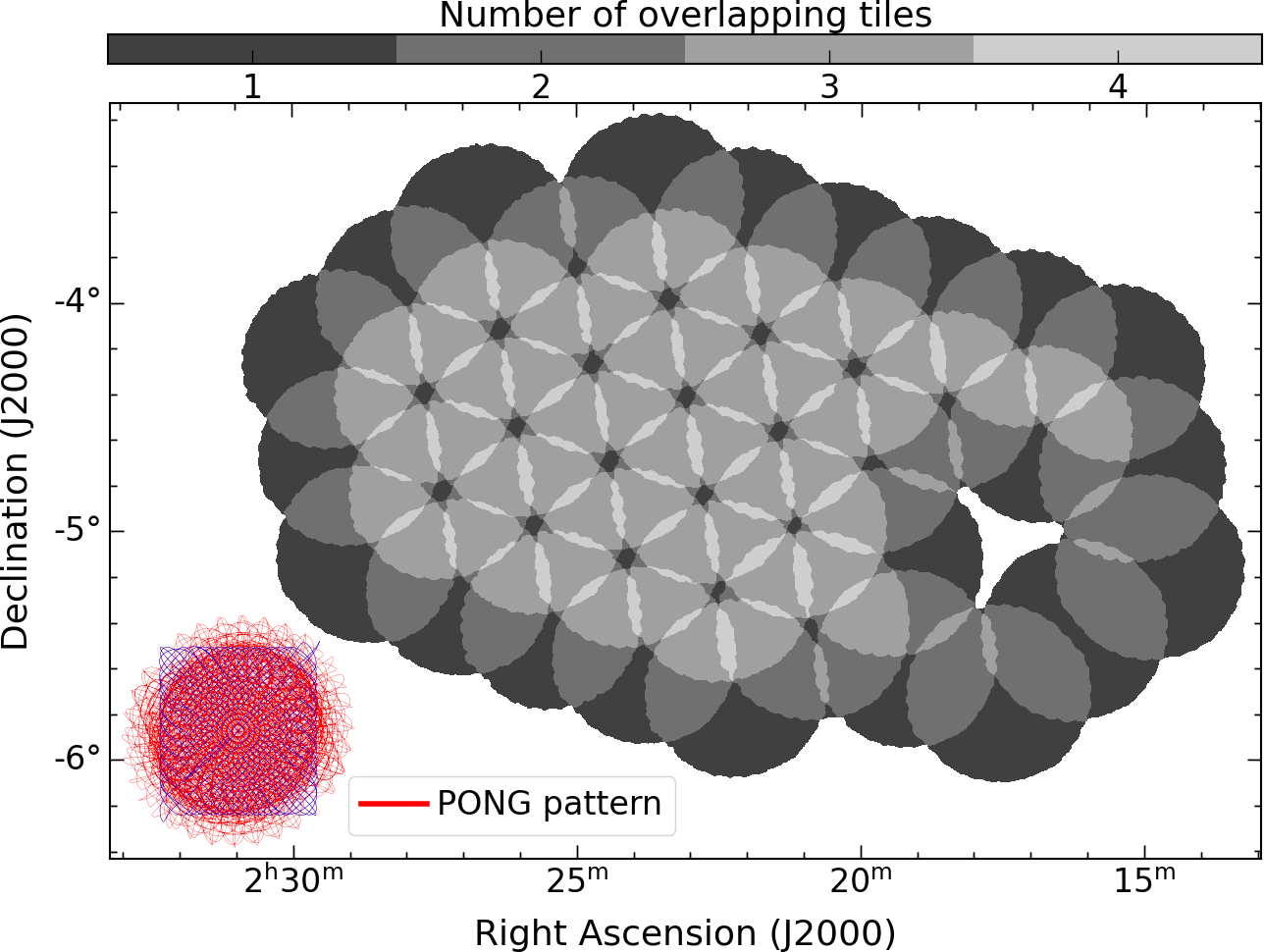}
    \caption{Layout of the S2LXS map showing the $42$ hexagonally arranged PONG tiles. An exposure time crop is applied to each individual PONG tile (see section \ref{section: data reduction} for details) before these are mosaicked. The colour map shows the number of PONG tiles that overlap across the map. The absence of data around $\alpha = 34^{\circ}.4542$, $\delta = -5^{\circ}.0986$ corresponds to the area of the deeper $1\sigma=1\,{\rm {mJy \,beam}^{-1}}$ S2CLS UKIDSS-UDS field observations \citep{Geach2017}. This region was not observed again in S$2$LXS.  In the lower left of the figure we also show an example of the telescope track for a PONG observation to illustrate the rotating PONG pattern \citep{Holland2013}.  The blue line shows the telescope track for a single rotation of the map, and the red line shows the complete PONG pattern (i.e., 8 rotations of the map).  
}
    \label{fig:layout}
\end{figure}

The S2LXS XMM-LSS  observations were conducted with SCUBA-2 \citep{Holland2013} on the James Clerk Maxwell telescope (JCMT) over $3$ years from July $2017$ to January $2020$, totalling $170$\,hours. The XMM-LSS field was mapped using the SCUBA-2 PONG mapping strategy \citep{Holland2013} for large fields in which the telescope array tracks across a target multiple times bouncing off the edges of a defined rectangular map area. Once the PONG pattern fills the map area the map is rotated and the pattern is repeated at the new angle. In Figure \ref{fig:layout} we show an example of the telescope track to illustrate the rotating PONG pattern. Observations were conducted for S2LXS XMM-LSS using the SCUBA-2 PONG $1800$\,arcsec diameter map pattern, which has a telescope scanning speed of $400\,{\rm arcsec\, s}^{-1}$ with $8$ rotations of the map during each observation to ensure a uniform coverage of the field. The final S2LXS XMM-LSS field map is a mosaic of $42$ hexagonally arranged $1800$\,arcsec diameter PONG tiles with each tile overlapping with its neighbours. We show the layout of the PONG tiles in Figure \ref{fig:layout}. There is a gap in the observations centred on $\alpha = 34^{\circ}.4542$, $\delta = -5^{\circ}.0986$. This absence of data reflects the area of the deeper ($1\sigma=1\,{\rm mJy\,beam}^{-1}$) S2CLS observations of the UKIRT Infrared Deep Sky Survey-Ultra Deep Survey field \citep[S2CLS UKIDSS-UDS;][]{Geach2017}. This region was not observed again in S$2$LXS XMM-LSS. 

\begin{figure}
	\includegraphics[width=\columnwidth]{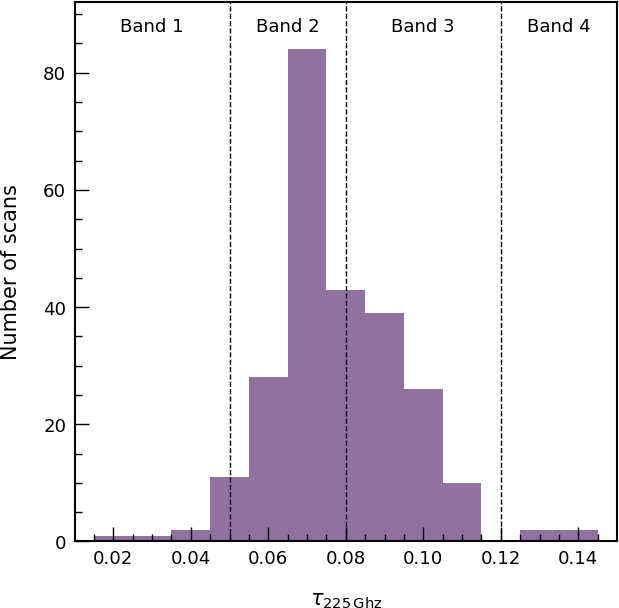}
    \caption{Zenith opacity at $225\,{\rm GHz}$. The opacity is recorded at the beginning and end of each PONG scan. We show the distribution of the average of these measurements (i.e., $\tau_{225}= (\tau_{\rm start} + \tau_{\rm end})/2$). For clarity we also show the boundaries of the JCMT weather bands.} 
    \label{fig:tau}
\end{figure}

The original objective of S2LXS was to map the XMM-LSS field to a depth of $1\sigma\simeq2\,{\rm mJy\,beam}^{-1}$ at $850\,\mu{\rm m}$. Progress of the survey was slow due to ongoing bad weather conditions on Mauna Kea and S2LXS XMM-LSS was consequently scaled back with the revised aim to map the S2LXS XMM-LSS field to a target sensitivity of $1\sigma\simeq4\,{\rm mJy\,beam}^{-1}$. At this depth the primary objective of S2LXS, to uncover the intrinsically rare, bright-end of the SMG population, is still realised. The individual PONG observations were limited to an integration time of $40\,$minutes to allow for accurate monitoring of variations in the observing conditions with regular pointing observations being made throughout the night. Each of the $42$ PONG tiles were observed over a minimum of $5$ scans (a minimum combined observing time of $3$ hours) to achieve the target sensitivity of $1\sigma\simeq4\,{\rm mJy\,beam}^{-1}$. The majority of observations ($97$ per cent) were conducted in band 1, 2 or 3 weather conditions (i.e., $\tau_{225\,{\rm GHz}}\leq0.12$). The remaining $3$ per cent of observations were taken in band 4 conditions in which the opacity at $225\,{\rm GHz}$ is in the range $0.12<\tau_{225\,{\rm GHz}}<0.16$ (see Figure \ref{fig:tau}) . In the following section we describe the process used to produce the S2LXS XMM-LSS maps. 

\subsection{Data reduction}\label{section: data reduction}

\begin{figure}
	\includegraphics[width=\columnwidth]{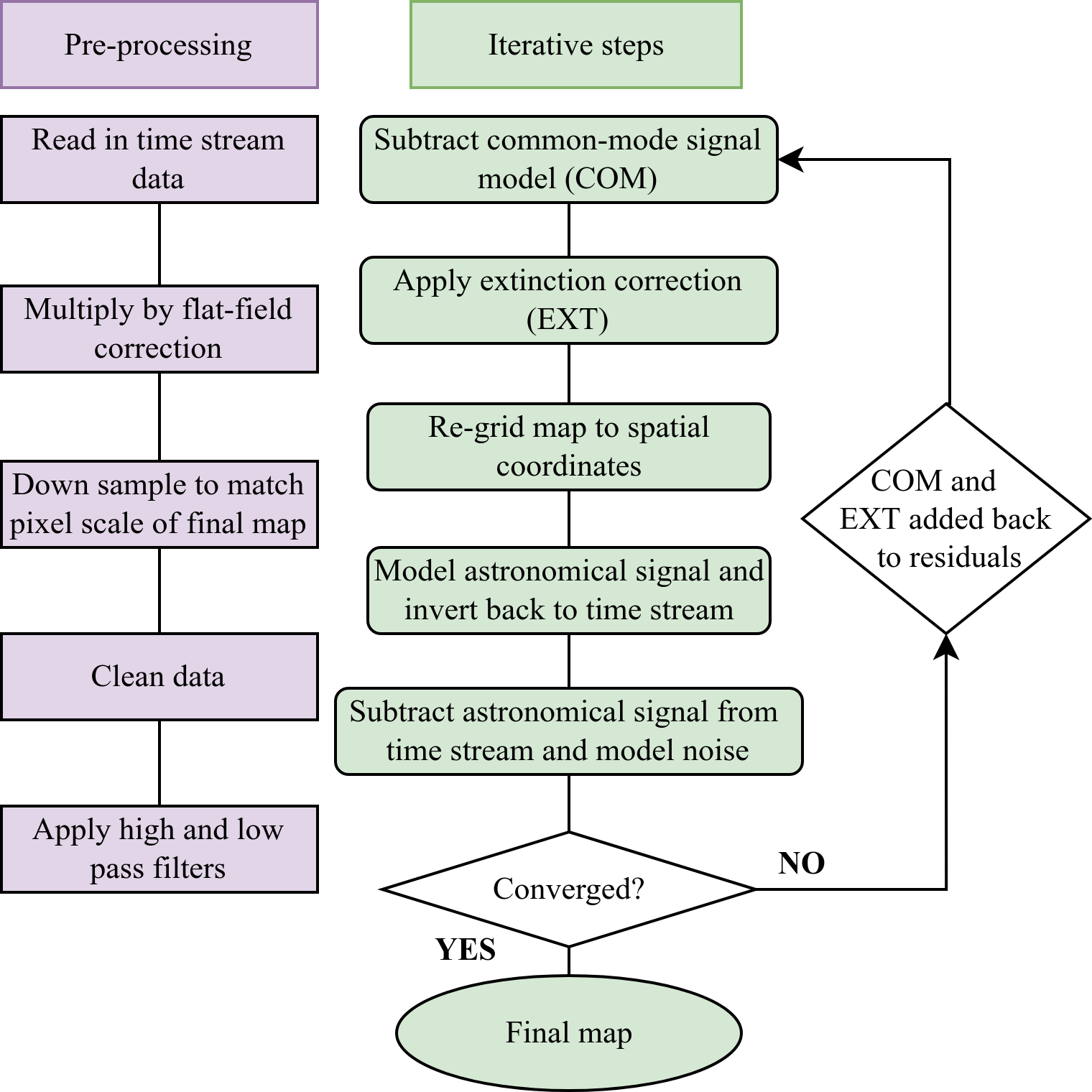}
    \caption{Flowchart illustrating the main stages of {\sc dimm} data reduction}
    \label{fig:dimm}
\end{figure}

The SCUBA-2 bolometers each record a time-varying signal containing contributions from astronomical signal, atmospheric extinction and noise. We reduce the S2LXS time-series data using the Dynamical Iterative Map-Maker ({\sc dimm}) tool, part of the Submillimetre Common User Reduction Facility software package \citep[{\sc smurf};][]{Chapin2013}. The primary purpose of this data reduction is to extract the astronomical signal from the SCUBA-2 bolometer time streams and to bin the resulting data into a two-dimensional celestial projection. {\sc dimm} includes a specialised `blank-field' configuration with parameters tuned for data reduction in extragalactic surveys. These parameters were optimised in the SCUBA-2 Cosmology Legacy Survey \citep[S2CLS;][]{Geach2017}. Taking this optimised configuration as the starting point we run several test data reductions using the S2LXS XMM-LSS raw data, varying the {\sc dimm} parameters and analysing the output (the reduced data). Our aim is to find the combination of parameters that best reduce residual noise in the S2LXS XMM-LSS map without overly compromising the astronomical signal. For full details of {\sc dimm} see \citet{Chapin2013}. Here we provide a short description of the main stages of {\sc dimm} data reduction (illustrated in the flow chart in Figure \ref{fig:dimm}) and describe the specific parameters we use to reduce the S2LXS XMM-LSS data.

The first step  in {\sc dimm} data reduction is the pre-processing stage in which the time-stream data are down-sampled and cleaned. The response of each bolometer to changing sky power is measured at the beginning and end of each observation (flat-field scans). In the pre-processing stage the raw data are first multiplied by a flat-field correction (estimated using the flat-field scans) to calibrate the bolometers. The time-streams are then down-sampled to a rate that matches the pixel scale of the final map ($2$\,arcsec). Next the time-stream data are cleaned for short-duration and high-amplitude spikes. In this step the signal from bright point sources can be mistaken for high-amplitude noise spikes and erroneously excluded. To avoid this we adopt a conservative threshold of $5\sigma$ and a box width of 50 time slices. At each time slice the median value of samples within a box centred on the time slice is measured. A time slice is flagged as a spike if the residual between the time slice value and this median value is greater than $5\times$ the local noise (the standard deviation of values in the neighbouring `down-stream' box). Next any sudden steps in the time series are identified and removed. Gaps are filled using a linear interpolation of the $50$ time slices preceding and the $50$ time slices following the excluded data. An order $1$ polynomial (i.e., linear) estimate of the base-line is then removed from each bolometer. In the final step of the {\sc dimm} pre-processing stage each bolometer time stream is independently filtered. A high pass filter is used to remove frequencies that correspond to spatial scales of $\theta>150$\,arcsec (to suppress large scale structures) and a low pass filter used to exclude data corresponding to angular scales of $\theta<2$\,arcsec (so that the beam is fully sampled).

Next {\sc dimm} begins an iterative process. First the common-mode signal (the average signal seen by all bolometers) is independently modelled for each SCUBA-2 sub-array and subtracted from the cleaned time-stream data. A multiplicative extinction correction (derived from atmospheric opacity measurements from the JCMT water vapour monitor) is then applied. In the second and any subsequent iterations the model of the astronomical signal estimated in the previous iteration is added back into the time series data at this stage. The next step is to model the astronomical signal. To do this the time series data are first binned onto a two-dimensional celestial projection. Each pixel in this two-dimensional grid is sampled many times by independent bolometers owing to the PONG scanning pattern and so, assuming that previous steps have removed all other sources of emission, the astronomical signal can be accurately estimated for a given pixel by taking the weighted average of the bolometer values that contribute to that pixel. The astronomical signal model is inverted back to the time stream and subtracted from the data, leaving just residual noise (made up of instrumental noise and atmospheric effects). This step in the iterative process can also be used to perform map-based despiking in which the scatter in the samples of a given map pixel is used to exclude outliers in the time-stream data. Since real astronomical sources have a fixed spatial location, bright point sources are unlikely to be falsely identified as spikes in this approach and so we can use a more stringent threshold here. Time-series residuals that differ by more than $3\sigma$ from the mean value in a map pixel are flagged and these are not used in following iterations. Finally a noise model is estimated for each bolometer by measuring the residual. This noise model is only estimated on the first iteration and is used to weight the data during the mapping process in subsequent iterations. This iterative process continues until either $20$ iterations have completed or the normalized mean change between consecutive maps is less than $0.05$. If further iterations are needed then all models (except the astronomical signal which is added back in later) are added back into the time-stream residuals.

To convert the map from units of $\rm{pW}$ to $\rm {Jy\,beam}^{-1}$ we apply a flux conversion factor (FCF) to the reduced map. We use the recently revised values for the standard FCF from \citet{Mairs2021}, and for observations taken before 30 June 2018 we adopt the standard FCF value of $516\,{\rm Jy\,beam}^{-1}\,{\rm pW}^{-1}$ and for observations taken after this date we use the standard FCF value of $495\,{\rm Jy\,beam}^{-1}\,{\rm pW}^{-1}$. We also apply an upwards correction of $10$ per cent \citep[estimated in S2CLS, see ][]{Geach2017} to account for the loss of flux density due to the filtering steps of the data reduction. The absolute flux calibration is estimated to be accurate to within $15$ per cent \citep[e.g.,][]{Geach2017}. 

Each of the approximately $40$\,minute PONG observations (PONG scans) are reduced independently to produce a set of individual maps for each of the $42$ PONG regions. We combine all scans for each PONG region using the {\sc picard} package within {\sc smurf} and the recipe {\sc mosaic\_jcmt\_images}. This recipe combines the maps using inverse-variance weighting with the variance of the final map calculated from the input variances. This gives us the $42$ PONG tiles which will make up the final S2LXS XMM-LSS map. 
 
\begin{figure}
	\includegraphics[width=\columnwidth]{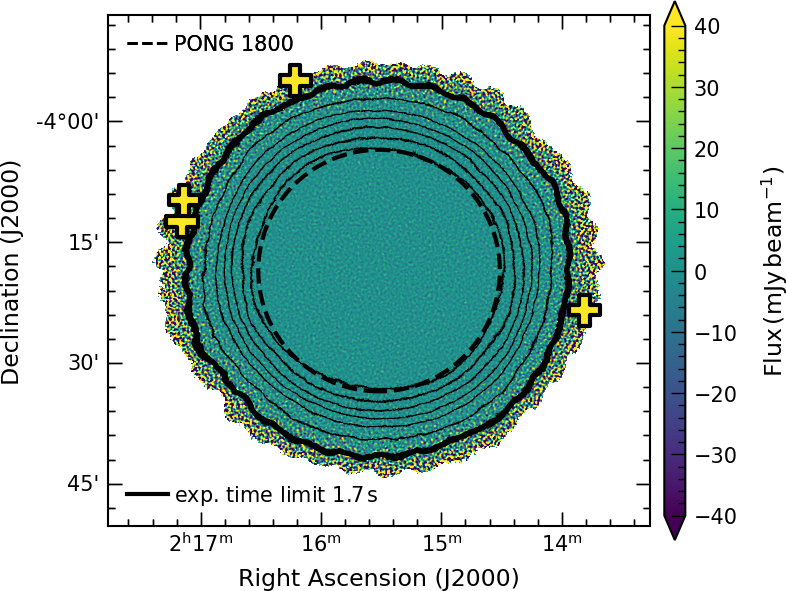}
    \caption{An example PONG tile (matched-filtered flux density map). The colour bar shows the flux density in mJy and the dashed circle shows the PONG map diameter of $1800$\,arcsec. Contours plotted in a thin black line are at $5$-$25\,{\rm s}$ in steps of $5\,{\rm s}$, and illustrate how the exposure time decreases in the over-scan region (outside the  diameter of the PONG). The optimum exposure time cut of $1.7{\rm s}$ is shown in the thick black line (see text for details of how this limit is derived and used). We also plot all $>5\sigma$ detections in this tile (crosses coloured with respect to the colour bar). All these (confirmed as spurious) detections lie in the over-scan region of the PONG tile with flux densities $>200\,{\rm mJy}$. These erroneous detections highlight the need to apply a crop to the over-scan region of the PONG tiles.}
    \label{fig:tile}
\end{figure}

The PONG mapping strategy provides uniform coverage within the PONG diameter (i.e., within a $1800$\,arcsec diameter). However, outside the PONG diameter (the over-scan region) each map pixel in the PONG scan is sampled less frequently. As the map pixels are sparsely sampled at the edges of the over-scan region the steps taken in the {\sc dimm} reduction to remove high amplitude spikes perform poorly. Consequently, at the edges of this over-scan region we see a multitude of very high (low) value pixels.  The paucity of the sampling also results in variance estimates which are not robust. This is problematic as a very high pixel value combined with an inaccurate variance estimate can masquerade as a bright point source in the outskirts of a PONG tile (see Figure \ref{fig:tile}). To mitigate this we apply an exposure time crop to the PONG tiles with the aim of eliminating these (obviously) spurious sources, whilst preserving as much of the data as possible. We iterate through increasingly stringent exposure time cuts until we find the minimum exposure time limit that effectively excludes these high (low) value pixels in the over-scan regions. In Figure \ref{fig:tile} we highlight the optimum exposure time limit of $1.7\,{\rm s}$.
 
We use the software package {\sc SWarp} \citep{Betrin2002}  to individually resample the PONG tiles (using nearest neighbour interpolation) to match the astrometric projection and pixel grid of the final science map. The resampled PONG tiles are then combined using inverse variance weighting to produce a map of the S2LXS XMM-LSS field (see Figure \ref{fig:layout} for the layout of the PONG tiles). 

To optimize the detection of point sources we apply a matched filter to the final mosaic using the {\sc picard} recipe {\sc scuba2\_matched\_filter}. Matched filtering consists of two stages. First the S2LXS map is smoothed with a Gaussian of full-width-half-maximum (FWHM) $30$\,arcsec and the result is subtracted from the original map. The purpose of this step is to remove any large-scale residual noise (i.e., on scales larger than the beam) not removed by the filtering applied during {\sc dimm} data reduction. The same smoothing operation and subtraction are applied to a Gaussian kernel of FWHM $14\,$arcsec (i.e., equal to the Airy disc radius at $850\,\mu{\rm m}$). In the second and final step of matched filtering the background-subtracted map is convolved with this kernel.

\begin{figure*}
	\includegraphics[width=1\linewidth]{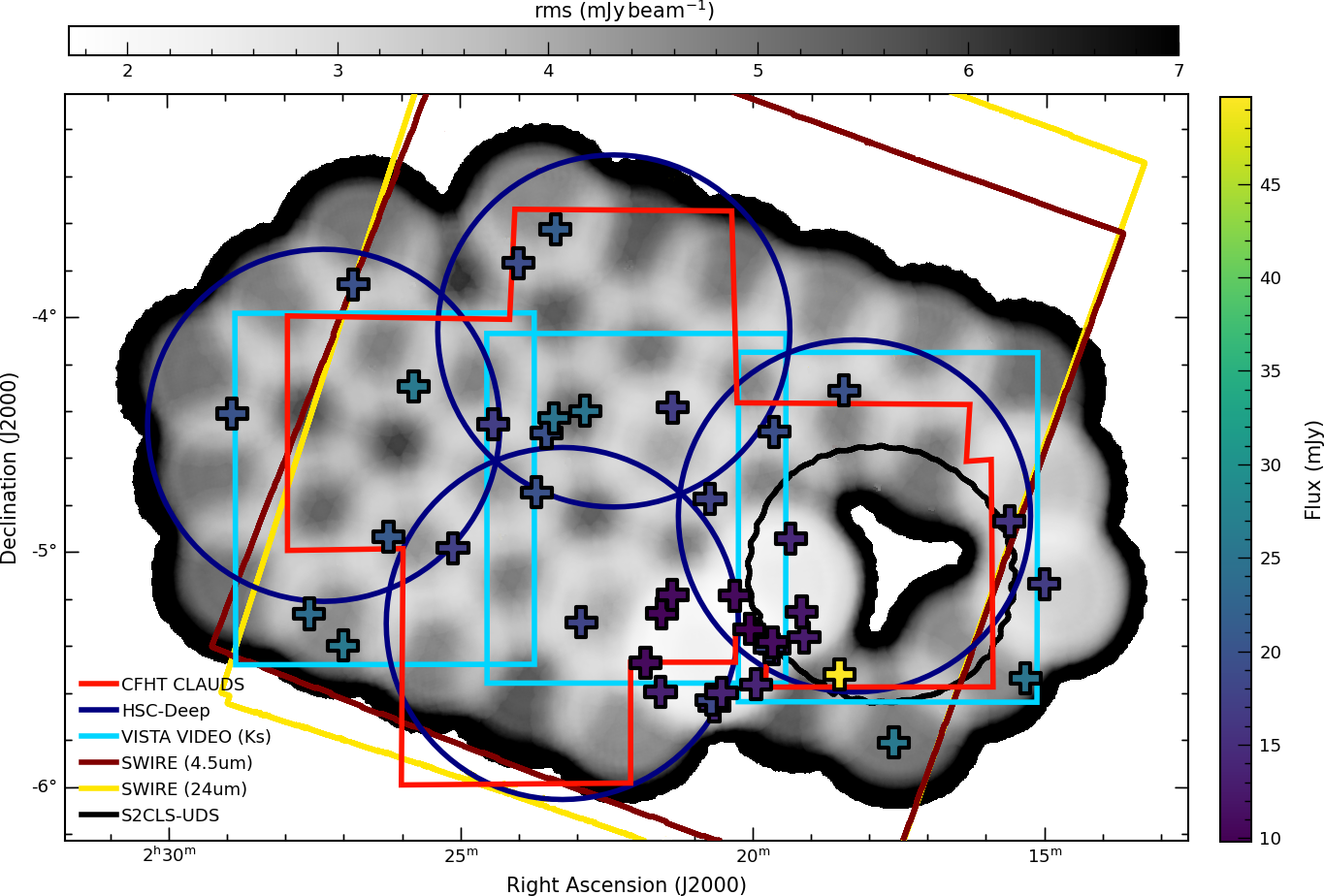}
    \caption{S2LXS XMM-LSS matched filtered instrumental noise map. The mosaic is made up of $42$ hexagonally arranged PONG tiles which are cropped to an exposure time of $1.7\,{\rm s}$. The colour map shows the $1\sigma$ depth in $\rm mJy \, beam^{-1}$ (horizontal colour bar). The S2LXS XMM-LSS map median rms is $1\sigma \simeq 4\,{\rm mJy\,beam}^{-1}$. The absence of data corresponds to the previously observed S2CLS UKIDSS-UDS field \citep[$\alpha = 34^{\circ}.4542$, $\delta = -5^{\circ}.0986$ -- black outline; ][]{Geach2017}, which was not observed in S$2$LXS. We overplot the footprints of CFHT CLAUDS ($u^*$), the HSC-Deep survey \citep{Aihara2018}, the VISTA-VIDEO survey \citep{Jarvis2013}, and SWIRE $4.5\,\mu{\rm m}$ and $24\,\mu{\rm m}$ imaging \citep{Lonsdale2003} to demonstrate the coverage of the multi-wavelength data. The area of the S2LXS XMM-LSS field is fully encompassed by HerMES \citep{Oliver2012}, the HSC-Wide survey \citep{Aihara2018} and XMM-LSS \citep{Pierre2004}, so the footprints of those surveys are not shown. The crosses in this figure indicate the positions of the $40$ ${>}5\sigma$ sources detected in the S2LXS XMM-LSS field, coloured with respect to the observed flux (vertical colour bar). We note that whilst some of the S2LXS XMM-LSS sources appear to cluster (potentially tracing large scale structure), the location of these sources is more likely a consequence of the variation in instrumental noise across the map, in part due to the overlap of the PONG tiles, with the majority of sources ($67.5$ per cent) detected in regions where at least $3$ tiles overlap. This is not unexpected given that map is more sensitive ($1\sigma \simeq 3.6\,{\rm mJy\,beam}^{-1}$) in these overlap regions.}  
    \label{fig:mosaic}
\end{figure*}

We show the S2LXS XMM-LSS field matched filtered instrumental noise mosaic in Figure \ref{fig:mosaic}. S2LXS did not re-observe areas covered in S2CLS, hence the absence of data in the region corresponding to the S2CLS UKIDSS-UDS field ($\alpha = 34^{\circ}.4542$, $\delta = -5^{\circ}.0986$). We overplot the footprints of CFHT CLAUDS ($u^*$), SWIRE mid-infrared ($24\,\mu{\rm m}$) and near-infrared ($4.5\,\mu{\rm m}$) imaging \citep{Lonsdale2003}, the HSC survey \citep[Deep tier;][]{Aihara2018} and the VISTA-VIDEO survey \citep{Jarvis2013}, highlighting the extent of multi-wavelength coverage for the S2LXS XMM-LSS field. We note that the area of the S2LXS XMM-LSS field is fully encompassed by HerMES \citep{Oliver2012}, the HSC-Wide survey \citep{Aihara2018} and XMM-LSS \citep{Pierre2004}, and so the footprints of these surveys are not plotted. The position of the $40$ ${>}5\sigma$ sources that make up the S2LXS XMM-LSS catalogue are also shown (see section \ref{section:source extraction} for details of the catalogue).

\subsection{Astrometric refinement and registration}

During JCMT observations standard calibrators are regularly observed to identify and correct for drifts in the telescope pointing. Typical corrections are of the order of $1$-$ 2$\,arcsec. To refine the astrometry of the S2LXS XMM-LSS map we require archival data from a survey which offers full coverage of the S2LXS XMM-LSS field. The HerMES and HSC-Wide survey both fully encompass the area of the S2LXS XMM-LSS science map. However, HSC-Wide is an optical survey and so samples a different population of galaxies compared to the population of dusty star-forming galaxies targeted in S2LXS XMM-LSS. In our tests (using the stacking method detailed below) with the HSC-Wide \textit{y}-band data ($\lambda_{\rm eff} = 0.9762\,\mu{\rm m}$) we do not get an adequate signal-to-noise in the final stack for a robust measurement of the offset in $\alpha$ and $\delta$. Therefore, to refine the astrometry of our final science map we make use of the HerMES data, specifically the release 4 (DR4) xID250 catalogue. which comprises about $70{,}000$ sources. Sources in the HerMES catalogue are blind detections at $250\,\mu{\rm m}$ with fluxes extracted in all \textit{Herschel} Spectral and Photometric Imaging Receiver (SPIRE) bands \citep{Roseboom2010,Roseboom2012,Oliver2012}. We exclude sources that are detected at the edge of the \textit{Herschel} SPIRE map, sources with ${S}_{250\mu {\rm m}}=0$ and  sources that fall outside the footprint of the S2LXS XMM-LSS mosaic, leaving us with a sample of approximately $30{,}000$ sources. The individual PONG observations were taken over a prolonged period of $3$ years and so we would expect the drift in the telescope pointing to be different for individual PONG scans. However, there are not a sufficient number of HerMES sources within the footprint of each PONG tile to robustly measure the astrometry offset for each observation, noting that when we tested this using the individual PONG scans the uncertainties in the offsets derived were bigger than the offset values themselves. Therefore, our aim here is to identify and correct for any large-scale offset in the S2LXS XMM-LSS map. To do this we centre $100$\,arcsec$\,\times$\,$100$\,arcsec cutouts of the S2LXS matched-filtered flux map at the positions of the HerMES sources and calculate the median value at each pixel position to create a median stack. We then use {\sc astropy \sc find\_peaks} (see section \ref{section:source extraction} for details of this function) to measure the centroid position of the peak signal in the stack. The offset in $\alpha$ and $\delta$ is taken as the difference between the coordinates of the peak signal and the coordinates of the centre of the stack. We repeat this process several times, each time updating the world coordinate system reference pixel coordinates with the small changes in $\alpha$ and $\delta$. The goal is to find the $\Delta\alpha$ and $\Delta\delta$ that maximize the signal-to-noise of the stack in the central pixel and so this process is repeated until both $\Delta\alpha$ and $\Delta\delta$ converge. We measure offsets of $\Delta \alpha \simeq +0.84''\pm 0.17''$ and $\Delta \delta \simeq -2.30''\pm 0.17''$ which are comparable to the expected magnitude of the pointing drift. We apply these small systematic corrections to the S2LXS science maps. 

\section{Analysis}
\subsection{Area Coverage}\label{section:area coverage}
\begin{figure}
	\includegraphics[width=\columnwidth]{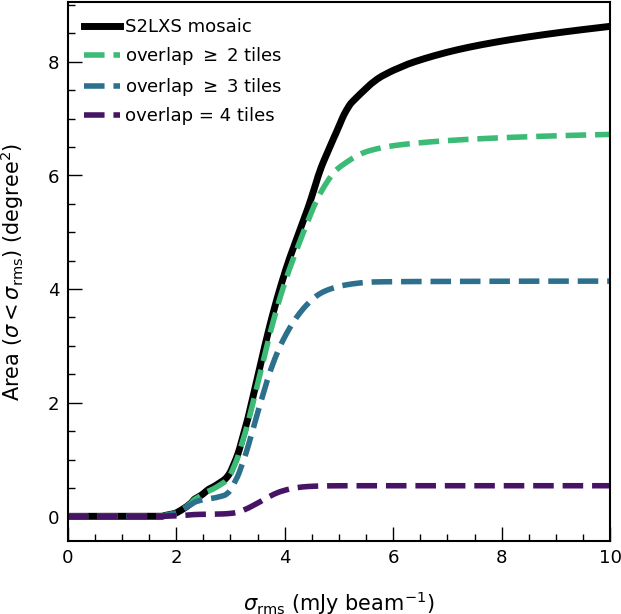}
    \caption{Cumulative area of the S2LXS XMM-LSS field mosaic as a function of instrumental noise (black solid line). We also show the cumulative area for regions of the map with $\geq2$, $\geq3$ and 4 overlapping PONG tiles, to highlight how the arrangement of the PONG tiles influences the depth of the S2LXS XMM-LSS map.}
    \label{fig:area}
\end{figure}

In Figure \ref{fig:mosaic} we show the S2LXS XMM-LSS instrumental noise map which has a median $1\sigma$ depth of $1\sigma\simeq4\,{\rm mJy\,beam}^{-1}$. The instrumental noise map is inhomogeneous, primarily due to (i) the variation in the median $1\sigma$ depth between the $42$ PONG tiles that make up the S2LXS XMM-LSS map (see section \ref{section: data reduction} for details of how the map was made), which ranges from $3.2$ to $7.7\,{\rm mJy\,beam}^{-1}$  and (ii) the arrangement of the PONG tiles in an overlapping hexagonal pattern. The difference in the median $1\sigma$ depth of individual PONG tiles is largely driven by the number of PONG scans that are stacked to produce each PONG tile. The majority of the PONG tiles were observed over $5$ PONG scans, however, due to the scheduling of observations $6$ PONG tiles were observed more than $5$ times, and so we have a larger number of PONG scans for these tiles (i.e., $6$, $7$, $10$, $10$, $18$, $21$ scans). The PONG tiles with additional scans are deeper (i.e., with $21$ scans we reach a median $1\sigma$ depth of $1\sigma=3.2\,{\rm mJy\,beam}^{-1}$) than the majority of PONG tiles that make up the map. The depth of the individual PONG tiles is also influenced by the weather conditions at the time of the observations and, in the case of one PONG tile (centred on $\alpha = 36.2566^{\circ}$, $\delta = -3.8432^{\circ}$) is due to one scan with only half the typical observation time ($20\,{\rm minutes}$). The instrumental noise across the S2LXS XMM-LSS map also varies depending on the number of overlapping PONG tiles. The median $1\sigma$ depth in the regions of the S2LXS XMM-LSS map with no overlapping PONG tiles is $1\sigma\simeq5.9\,{\rm mJy\,beam}^{-1}$, compared to $1\sigma\simeq 4.3\,{\rm mJy\,beam}^{-1}$ for regions with $2$ overlapping tiles and $1\sigma\simeq 3.6\,{\rm mJy\,beam}^{-1}$ for regions with $\geq3$ overlapping tiles. The instrumental noise also increases rapidly (to a maximum noise of $1\sigma_{\rm max}=27.0\,{\rm mJy\,beam}^{-1}$) in  the outer edges of the S2LXS XMM-LSS mosaic, where the over-scan regions of individual PONG tiles do not overlap with neighbouring tiles. In Figure \ref{fig:area} we show the cumulative area of the map as a function of depth highlighting the contribution due to the number of overlapping PONG tiles. The S2LXS XMM-LSS science map has an approximate area coverage of $9\,{\rm deg}^{2}$. 

The confusion limit ($\sigma_{\rm c}$; Scheuer 1957) is the flux level at which pixel-to-pixel variance ($\sigma^{2}$) ceases to reduce even with increased exposure time, due to the crowding of faint sources within the telescope beam. The confusion limit of an image is typically reached when the surface density of sources exceeds $0.03$ per beam \citep[e.g.,][]{Condon1974,Hogg2001}. The confusion limit is measured in previous SCUBA-2 surveys and estimated to be $\sigma_{\rm c}\simeq1\,{\rm mJy\,beam}^{-1}$ at $850\,\mu{\rm m}$ \citep[e.g.,][]{Geach2017,Simpson2019}. With a median $1\sigma$ depth of $1\sigma\simeq4\,{\rm mJy\,beam}^{-1}$ the variance in our maps is dominated by instrumental noise (even in deeper overlap areas) and so we do not consider the sub-dominant effect of confusion noise on the S2LXS XMM-LSS survey any further. 

\subsection{Source extraction}\label{section:source extraction}
The S2LXS XMM-LSS field has extensive multi-wavelength coverage that might ordinarily encourage a prior-based source extraction approach. However, about $20$ per cent of SMGs that are detected at $850\,\mu{\rm m}$ are undetected in sensitive near-infrared imaging at $K\gtrsim25.7\,{\rm mag_{\rm AB}}$ \citep[e.g.,][]{Dudzevi19}. To avoid overlooking this otherwise obscure minority of the SMG population we opt for a blind source extraction approach, in keeping with S2CLS \citep{Geach2017} and S2COSMOS \citep{Simpson2019}. We caveat that a limitation of using a blind source extraction approach is that we will not separate out multiple discrete sources blended in the coarse $15$\,arcsec resolution of the SCUBA-2 beam \citep[e.g.,][]{Simpson2015,Stach2018}. 

The S2LXS XMM-LSS maps have already been optimized for the detection of point source emission with the application of a matched-filter (see section \ref{section: data reduction} for details). We use {\sc Photutils}\footnote{In this work we use {\sc astropy} version $4$.$2$.$1$ and {\sc photutils} version $1$.$0$.$1$}, an {\sc astropy} affiliated python package to detect sources in the S2LXS XMM-LSS signal-to-noise ratio matched-filtered map. This package includes a function {\sc find\_peaks} to locate local maxima in an astronomical image that are above a specified threshold value. We set a detection value of $3.5 \sigma$. Setting a low detection threshold allows us to explore the statistical properties of sources detected at the lowest significance in the S2LXS map, noting that it is simple to execute further cuts  directly on the source catalogue. We also impose a separation limit, requiring sources to be separated by  ${>}15$\,arcsec (equivalent to the FWHM of the SCUBA-2 beam). If more than one source is detected above the detection threshold in a box size of $15$\,arcsec only the coordinates of the source with the highest signal-to-noise ratio are returned. This prevents double counting of bright sources if more than one pixel associated with a source exceeds the detection threshold. Whilst the bright-end of the $850\,\mu{\rm m}$ number counts is poorly constrained, these sources are expected to be rare \citep[e.g.,][]{bethermin2017,cowley2015}, with approximately $10$ sources with flux densities above $10\,{\rm mJy}$ per square degree \citep[e.g.,][]{Geach2017}. Assuming sources are randomly distributed the probability of finding two $10\,{\rm mJy}$ sources separated by ${<}15$\,arcsec is less than $0.1$ per cent, so we are unlikely to overlook any bright sources by imposing this separation limit. However, we note that if bright sources are clustered \citep[as seen at fainter fluxes e.g.,][]{Cairns2022,Greenslade2020} or are part of merging systems, imposing this separation limit may potentially exclude some sources. To mitigate this we can check for additional bright sources by visually inspecting the S2LXS XMM-LSS science map at the position of the sources in the survey catalogue, searching for elongated sources or clearly separated $>5\sigma$ signal-to-noise ratio peaks within the SCUBA-2 beam (noting that confirmation of multiple galaxies within the SCUBA-2 beam requires follow up by high resolution interferometers). The {\sc find\_peaks} function allows the user to input a user-defined centroid function to calculate the coordinates of a source to sub-pixel accuracy. We create a function that calculates the centroid by fitting a model of the SCUBA-2 instrumental point spread function\footnote{The SCUBA-2 point spread function (PSF) is well described by the superposition of two Gaussian functions: the primary component has a FWHM $13$\,arcsec and contains $98$ per cent of the total flux; and the secondary component has a FWHM $48$\,arcsec and contains $2$ per cent of the total flux \citep{Dempsey2013}.} \citep{Dempsey2013} to the 2-dimensional distribution of the data. Using {\sc find\_peaks} and the parameters above we detect $1966$ sources at ${>}3.5\sigma$ and $40$ at ${>}5\sigma$. We inspect the S2LXS XMM-LSS images at the positions of sources in the ${>}5\sigma$ catalogue to check for multiple bright sources clustered within the SCUBA-2 beam, but find no evidence for this.  The formal detection limit set for S2LXS XMM-LSS is $5\sigma$ and this is the threshold limit at which we define the survey catalogue. At this threshold we estimate that the false detection rate is $10$ per cent (see section \ref{section: false_detection} for details). We note that whilst some of the S2LXS XMM-LSS sources appear to cluster (potentially tracing large scale structure), the location of these sources in the S2LXS XMM-LSS map is more likely a consequence of the variation in instrumental noise across the map, in part due to the overlap of the PONG tiles, with the majority of sources ($95$ per cent) in the survey catalogue detected in deeper regions where ${\geq}2$ PONG tiles overlap. 

\subsection{Simulated SCUBA-2 maps}\label{section:simulated maps}

To test the reliability of our source extraction we create simulated SCUBA-2 maps projected onto the same pixel grid as the observed S2LXS XMM-LSS mosaic. To create our fake maps we must first produce a realistic instrumental noise map. We do this by creating a jackknife map, which are typically produced by randomly inverting the flux densities for $50$ per cent of the observed data before combining. This removes the astronomical signal from the stack and creates a realistic realisation of the instrumental noise. The S2LXS XMM-LSS mosaic is made up of $42$ PONG tiles that are produced by stacking the $\sim40$ minute PONG observations (i.e., the PONG scans, see section \ref{fig:dimm} for details). For each PONG tile the simplest way to invert $50$ per cent of the observed data is to invert the flux densities of half the PONG scans before combining. We do this for all PONG tiles composed of an even number of observations. However, the majority of the PONG tiles are observed over $5$ scans and so for these tiles we cannot simply invert the flux densities of half the PONG scans. Therefore, for these PONG tiles, we randomly select one PONG scan and reduce the data (using the dimm reduction process and parameters described in section \ref{fig:dimm}) in two consecutive halves, producing two reduced maps each composed of $50$ per cent of the data for the observation. For a PONG tile composed of $5$ scans (for example) we can then invert $50$ per cent of the observed data by inverting the flux densities of $2$ randomly selected scans and of one of these `split' scans (i.e., inverting the observed data for $2.5$ scans). 

\begin{figure} 
	\includegraphics[width=\columnwidth]{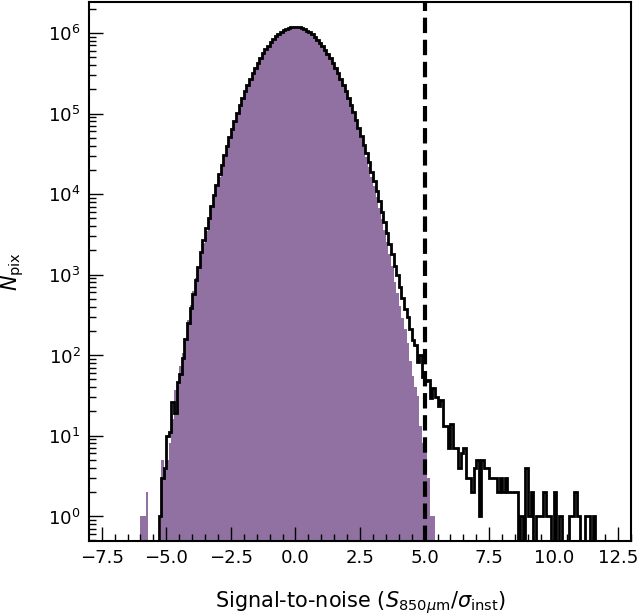}
    \caption{Distribution of pixel values in the matched-filtered signal-to-noise ratio jackknife map (filled histogram). We also show the distribution of pixel values from the S2LXS XMM-LSS matched-filtered signal-to-noise ratio map (black stepped histogram) and the S2XLS XMM-LSS formal detection threshold of $5\sigma$ (black dashed line). This figure illustrates the characteristic tail of astronomical signal at positive fluxes. We also see a small number of pixel values with a signal-to-noise of $<-5$ in the jackknife map. This is likely caused by isolated pixels with very low flux values in the sparsely sampled over-scan regions of the PONG maps.}
    \label{fig:jackknife_distribution}
\end{figure}

To create a corresponding jackknife tile for each PONG tile we invert the flux densities for $50$ per cent of the observed data (as described above) and then combine with the remaining observations using the {\sc picard} recipe {\sc mosaic\_jcmt\_images}. To produce the jackknife mosaic we then combine the $42$ jackknife tiles following the same method we used to create the S2LXS XMM-LSS mosaic (see section \ref{section: data reduction}). In brief, we first apply a $1.7\,{\rm s}$ exposure time crop to each of the jackknife tiles and then resample these individually to match the astrometric projection of the final science map. We then use inverse-variance weighting to combine the $42$ re-sampled jackknife tiles to produce the jackknife mosaic. In Figure \ref{fig:jackknife_distribution} we compare the distribution of pixel values for the  jackknife signal-to-noise ratio matched-filtered mosaic to the distribution of pixel values in the S2LXS XMM-LSS signal-to-noise ratio map. We also show the S2LXS XMM-LSS $>5\sigma$ detection limit. This figure highlights the characteristic tail of astronomical signal, which is pronounced at $>5\sigma$.    

The next step to create our simulated maps is to insert fake sources matching a realistic source count model into the jackknife noise map.  The faint end of the $850\,\mu{\rm m}$ number counts is well constrained and so we use the \citet{Schechter1976} fit from \citet{Geach2017} to model the source counts below $15\,{\rm mJy}$. However, the bright-end of the $850\,\mu{\rm m}$ number counts is not well constrained, with only $5$ extragalactic sources in S2CLS with $S_{850\,\mu{\rm m}}>15\,{\rm mJy}$.  Therefore, to model the source counts above $15\,{\rm  mJy}$ we fit a simple powerlaw to the S2LXS XMM-LSS observed number counts.

We simulate an observed SCUBA-2 source by convolving a point source of a given flux with the SCUBA-2 instrumental PSF \citep{Dempsey2013}, noting we impose a minimum flux limit of $1\,{\rm mJy}$ \citep[in keeping with][]{Geach2017}. The simulated SCUBA-2 sources are then injected at random positions in the jackknife noise map (we do not apply any conditions on the positioning of sources, these are inserted purely at random) to generate a simulated map.  We then use the {\sc picard} recipe {\sc scuba2\_matched\_filter} to apply a matched-filter to the simulated map and {\sc astropy find\_peaks} to detect sources in the map above the floor detection threshold value (${>}3.5$\,sigma), mimicking the source extraction method used on the S2LXS XMM-LSS mosaic (see section \ref{section:source extraction}).  A fake source is recovered if it is found above the detection threshold and at a maximum separation of $11$\,arcsec ($0.75\times$ the FWHM of the SCUBA-2 beam) from the input coordinates. We note that this matching threshold is somewhat arbitrary, but is intentionally generous \citep[and in keeping with the matching threshold used in][]{Geach2017,Simpson2019}. If there are multiple detections within this radius then the closest match is taken as the recovered source.  If a source is recovered then the fake source catalogue will include the recovered flux density, rms and coordinates of the source, as well as the signal-to-noise ratio of the detection.  To allow us to fully sample the flux regime we are interested in (i.e., $S_{850\,\mu{\rm m}}>15\,{\rm mJy}$) we repeat this process $100,000$ times, creating $100,000$ simulated maps and corresponding fake source catalogues.  For the source extraction on our simulated maps we use a floor detection limit of ${>}3.5\sigma$ to investigate the statistical properties of the simulated sources detected at low significance, and to compare to sources detected at the same threshold in the S2LXS XMM-LSS mosaic. However, for the remainder of this paper we assume the formal detection limit of S2LXS XMM-LSS which is $5\sigma$. 

\subsection{Flux boosting}\label{section: flux boosting}

\begin{figure}
	\includegraphics[width=\columnwidth]{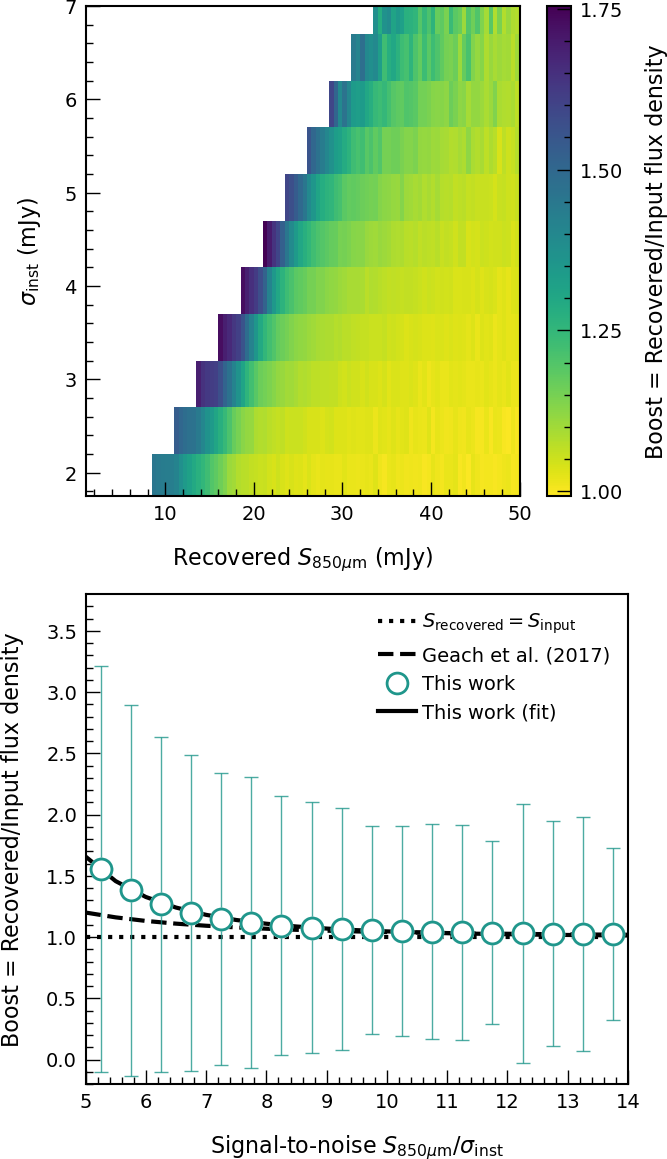}
    \caption{Flux boosting is measured by comparing the recovered flux density to the injected flux density for each source in the fake source catalogue. The top panel shows average flux boosting as a function of recovered (i.e., observed) flux density and local instrumental noise. In the bottom panel we show average flux boosting (uncertainties shown as $\pm1\sigma$) as a function of recovered signal-to-noise ratio (teal circles). We also plot the power law relation from S2CLS \citep[][dashed curve]{Geach2017} and the power law fit from this work (solid curve, see equation \ref{eq: flux_boost}). The dotted line shows $S_{\rm recovered}=S_{\rm input}$. The average effect of flux boosting is stronger in S2LXS XMM-LSS compared to S2CLS, which is likely motivated by the higher instrumental noise of the S2LXS XMM-LSS survey. As expected the effect of flux boosting is most significant for sources detected at relatively low signal-to-noise ratio.}
    \label{fig:flux_boosting}
\end{figure}

The flux density of galaxies detected at a relatively low signal-to-noise ratio is likely to be boosted upwards due to fluctuations in the noise \citep[e.g.,][]{Hogg1998}. We can estimate the effect of flux boosting on the S2LXS XMM-LSS survey by comparing the recovered flux density to the intrinsic flux density for each source detected at ${>}5\sigma$ in the fake source catalogue. In Figure \ref{fig:flux_boosting} we show average flux boosting as a function of observed flux density and local instrumental noise, and as a function of signal-to-noise ratio. As expected the effect of flux boosting is most significant at a relatively low signal-to-noise ratio. At $5\sigma$ the observed flux density is approximately $66$ per cent higher on average than the intrinsic flux density. We find that the relation between boosting ($\mathcal{B}$ = $S_{\rm obs}$/$S_{\rm true}$) and signal-to-noise ratio is well described by a power law (equation \ref{eq: flux_boost}). In keeping with \citet{Geach2017} we use a power law of the form ${\rm y} = 1 + a{\rm x}^{k}$, where $a$ and $k$ are constants, ${\rm y}=\mathcal{B}$ and ${\rm x}=({\rm SNR}/5)$:  

\begin{ceqn}
\begin{align}
\mathcal{B} = 1+0.66 \left(\frac{{\rm SNR}}{5}\right)^{-3.83} 
\label{eq: flux_boost}
\end{align}
\end{ceqn}

\noindent In Figure \ref{fig:flux_boosting} we compare the average effect of flux boosting in S2LXS XMM-LSS to S2CLS \citep{Geach2017}, and find that the average effect of flux boosting is stronger in S2LXS XMM-LSS. We attribute this difference to the higher instrumental noise of the S2LXS XMM-LSS map.

\begin{figure*}
	\includegraphics[width=1\linewidth]{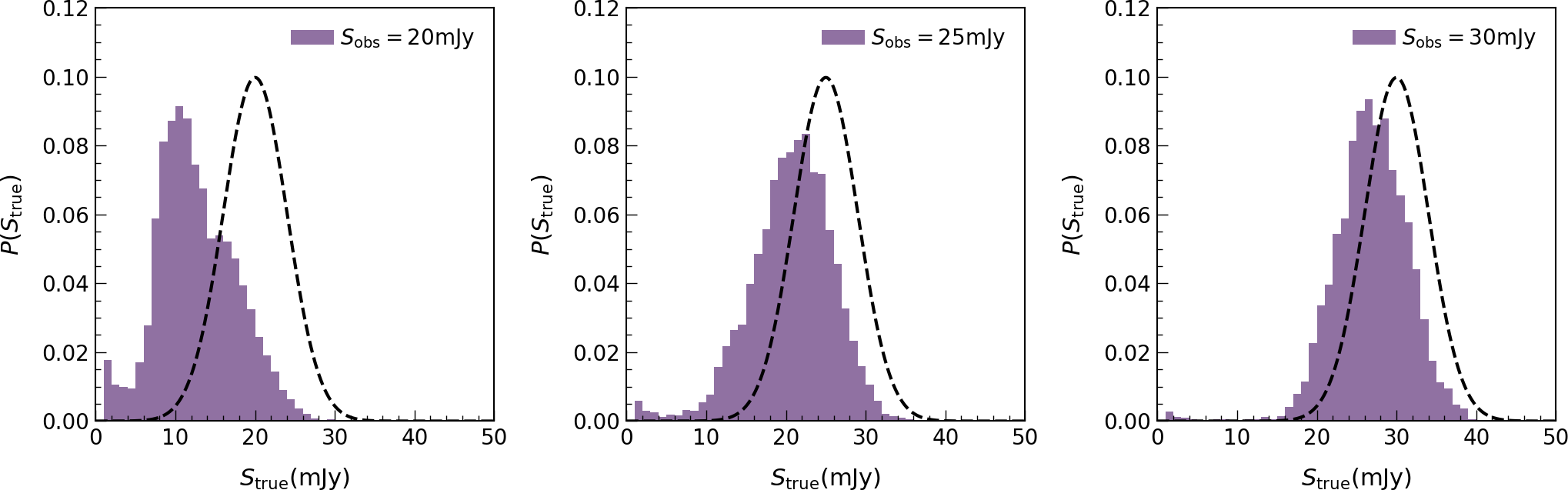}
    \caption{Distribution of injected (true) flux densities of sources in the fake source catalogue in bins of $S_{\rm obs}$ (shown in legend) and $\sigma_{\rm inst}=4\,{\rm mJy}$ (approximately the $1\sigma$ median map depth). We also show a normal distribution centred on $S_{\rm obs}$ with a standard deviation of $4\,{\rm mJy}$ (dashed line). As expected for sources that are observationally bright (i.e., observed at higher signal-to-noise ratio with $\sigma_{\rm inst}$ fixed) the effect of flux boosting is diminished.}
    \label{fig:flux_boost_dist}
\end{figure*}

Due to the statistical nature of flux boosting the intrinsic (true) flux density of a source with an observed flux density $S_{\rm obs}$ is drawn from a distribution \textit{p}\,($S_{\rm true}$). We can empirically estimate the effect of flux boosting by measuring the histogram of injected flux densities of fake sources in bins of observed flux density and instrumental noise (see Figure \ref{fig:flux_boost_dist} for example). This empirical approach has been shown to be comparable \citep[e.g.,][]{Geach2017} to traditional Bayesian techniques \citep[e.g.,][]{Jauncey1968,Coppin2005}. To estimate the deboosted (true) flux density for each source in our ${>}5\sigma$ catalogue we draw $1000$ samples from the \textit{p}\,($S_{\rm true}$) distribution measured from our simulations for an observed flux density and local instrumental noise corresponding to that of the real source. We take the mean of the samples as the deboosted flux density of the real source and the standard deviation of the sample as the error on the deboosted flux density.

 \subsection{Completeness}\label{section:completeness}

\begin{figure}
    \centering
	\includegraphics[width=0.94\columnwidth]{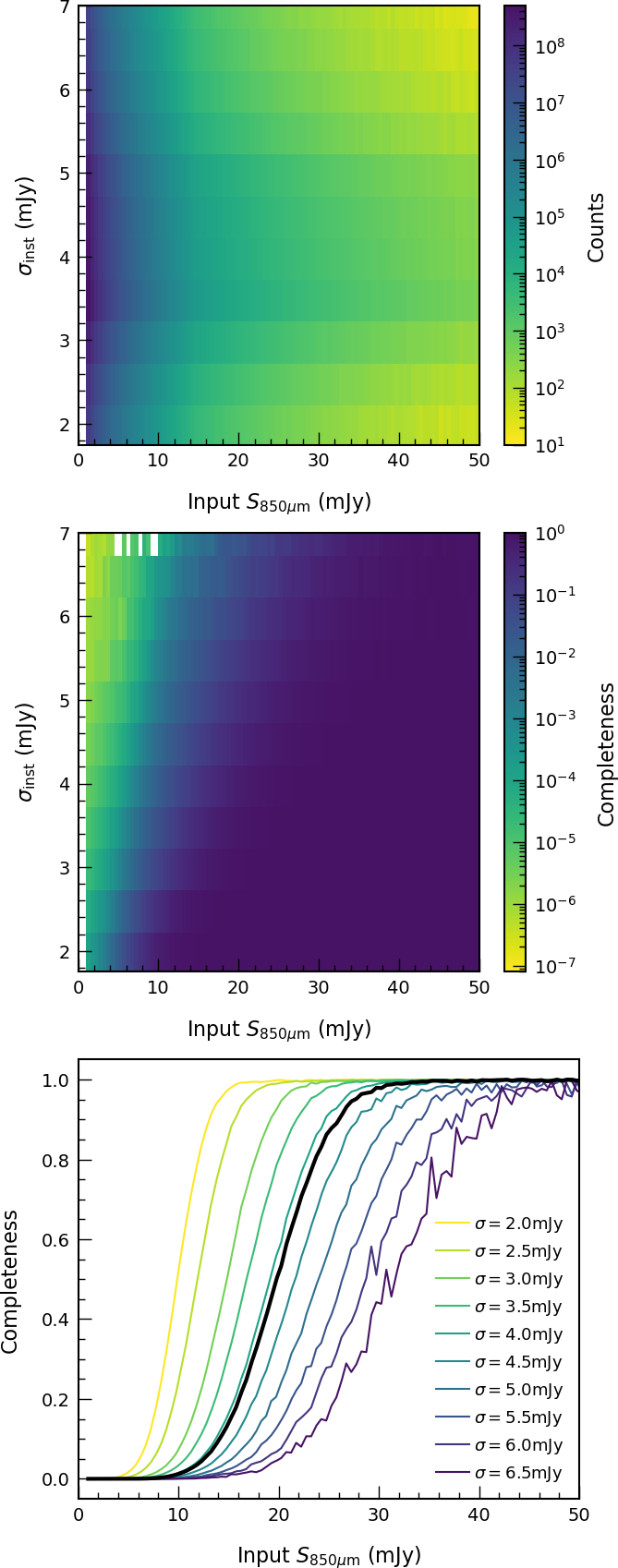}
    \caption{S2LXS XMM-LSS survey completeness estimated by comparing the number of sources injected into our fake maps to the number recovered at a detection threshold of $5\sigma$. In the top panel we show the injected fake source counts and in the centre panel the completeness ratio, both as a function of input flux density and local instrumental noise. In the bottom panel we plot the survey completeness at the median map depth ($\sigma_{\rm inst}\simeq 4\,{\rm mJy\,beam}^{-1}$, black solid line) and at a range of map depths (see legend) to show the variation in completeness across the full S2LXS XMM-LSS survey area. }
    \label{fig:complete}
\end{figure}

We investigate the completeness of the S2LXS XMM-LSS survey by comparing the number of sources injected into the simulated SCUBA-2 maps to the number of sources recovered at ${>}5\sigma$, evaluated in bins of input flux density and local instrumental noise. In Figure \ref{fig:complete} we show the injected source counts and the estimated S2LXS XMM-LSS survey completeness as a function of input (true) flux density and local instrumental noise. The survey $50$ per cent ($90$ per cent) completeness at the median map depth ($1\sigma\simeq 4\,{\rm mJy\,beam}^{-1}$) is $19.8\,{\rm mJy}$ ($26.0\,{\rm mJy}$). In Figure \ref{fig:complete} we plot the survey completeness across a range of map depths ($2.0$ to $6.5\,{\rm mJy}$) to demonstrate the variation in survey completeness over the full survey area. Our binned completeness values act as a lookup table and we use $2$--dimensional spline interpolation to estimate the completeness rate at the de-boosted flux density and local instrumental noise for each source in the S2LXS XMM-LSS ${>}5\sigma$ catalogue. 

\subsection{Positional uncertainty}
\begin{figure}
	\includegraphics[width=\columnwidth]{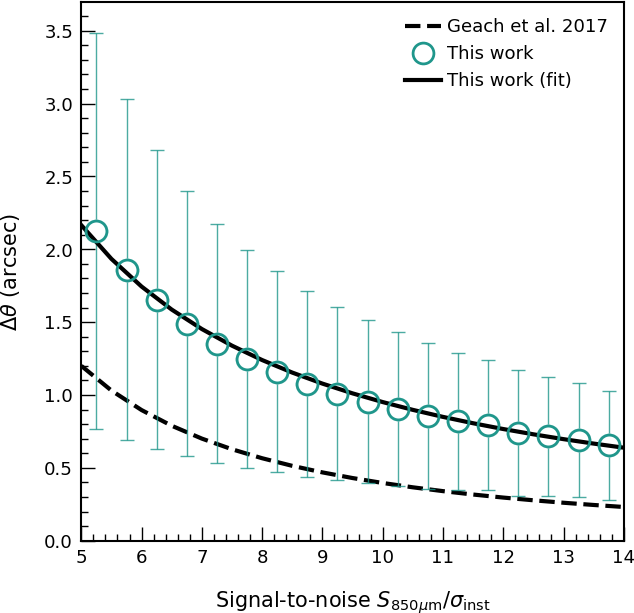}
    \caption{Positional offset $(\Delta \theta)$ estimated by calculating the difference between the injected and recovered coordinates for each source in the fake catalogue. The average positional offset is shown (with $\pm1\sigma$ uncertanties) as a function of signal-to-noise ratio (teal circles). We plot the power law relation from \citet{Geach2017} (dashed line) and the power law fit from this work (solid line, see equation \ref{eq: pos_uncert}).  The positional offset is almost double the average positional uncertainty found in S2CLS, possibly due to S2CLS being a much deeper survey ($1\sigma\simeq1\,{\rm mJy}$).}
    \label{fig:pos_uncert}
\end{figure}

Positional uncertainties are expected to scale with signal-to-noise ratio for a given Gaussian-like beam \citep{Condon1997,Ivison2007}. We measure the positional offset ($\Delta \theta)$ for each source in our fake catalogue by calculating the difference between the injected and recovered coordinates. In Figure \ref{fig:pos_uncert} we show the average positional offset as a function of signal-to-noise ratio for sources in the fake catalogue recovered at ${>}5\sigma$. We find that the relation is well described by a simple power law of the form

 \begin{ceqn}
\begin{align}
\Delta \theta = 2.17\,{\rm arcsec} \times \left(\frac{{\rm SNR}}{5}\right)^{-1.19}. 
\label{eq: pos_uncert}
\end{align}
\end{ceqn}

The average positional offset for sources detected at $5\sigma$ in S2LXS XMM-LSS is approximately $2$\,arcsec.  This offset is consistent with theoretical predictions \citep[e.g., equation B8 of][]{Ivison2007}, and the results of some SCUBA-2 wide-field surveys \citep[e.g.,][]{shim2020,Simpson2019}.  However, this is almost double the average positional uncertainty found in S2CLS (see Figure \ref{fig:pos_uncert}), possibly due to S2CLS being a much deeper survey ($1\sigma\simeq1\,{\rm mJy}$). The positional offset (equation \ref{eq: pos_uncert}) should be taken into account when identifying multi-wavelength counterparts to sources in the S2LXS XMM-LSS ${>}5\sigma$ catalogue.
 
\subsection{False detection rate}\label{section: false_detection}

\begin{figure}
\includegraphics[width=\columnwidth]{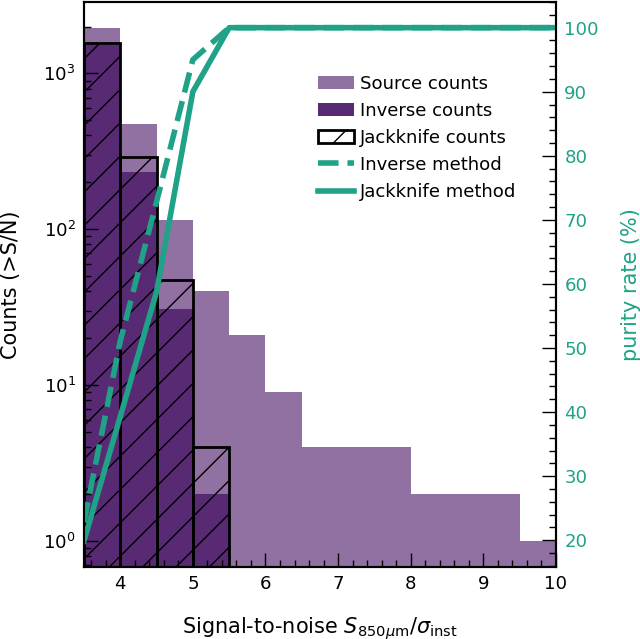}
    \caption{Purity rate estimated by comparing the number of sources in the S2XLS XMM-LSS science map to the number of spurious detections in the jackknife map (solid line) above a signal-to-noise threshold. The purity rate is also estimated by comparing to the `source' counts in the inverted S2XLS XMM-LSS science map (dashed line). We show the counts obtained from the S2XLS XMM-LSS science map (light purple), the inverse map (dark purple) and the jackknife map (hatched histogram).}
    \label{fig:false_detect}
\end{figure}

To estimate the false detection rate for the S2LXS XMM-LSS survey we compare the number of sources detected above a given threshold in the S2LXS XMM-LSS match-filtered signal-to-noise ratio map to the number of `sources' (i.e., spurious detections) detected above the same threshold in the matched-filtered jackknife map (see section \ref{section:simulated maps} for details of how the jackknife map is created). In Figure \ref{fig:false_detect} we show the number of sources detected in the S2LXS XMM-LSS science map compared to the number of sources detected in the jackknife map as a function of signal-to-noise ratio. We plot the purity rate ($P$) calculated as $P=(N_{\rm p} - N_{\rm n})/N_{\rm p}$, where $N_{\rm p}$ is the number of real+spurious sources detected above a given threshold in the  S2LXS XMM-LSS mosaic and $N_{\rm n}$ is the number of jackknife `source' detections above the same limit. We find a false detection rate (purity rate) of $10.0$ per cent ($90.0$ per cent) at ${>}5\sigma$ and $41.23$ per cent ($58.77$ per cent) at ${>}4.5\sigma$. 
 
As a check we repeat our analysis but instead of using the jackknife map to estimate the number of spurious detections, we use the inverse of the  S2LXS XMM-LSS match-filtered signal-to-noise map. The match-filtering process creates negative bowling around bright sources, and so we mask areas of negative bowling in the inverse map to avoid this contaminating our results. The estimated purity rate using this inverse method is comparable to that obtained using the jackknife approach. We show the results from both methods in Figure \ref{fig:false_detect}. In keeping with recent SCUBA-2 wide-field surveys \citep[e.g.,][]{Geach2017,Simpson2019} we use the false detection rate estimates from the jackknife approach for the remainder of this work. The S2LXS XMM-LSS survey formal detection threshold of $>5\sigma$ is chosen to give a $90$ per cent purity rate. At this purity rate we expect ${\sim}4$ of the $40$ sources in the ${>}5\sigma$ catalogue to be spurious.  We note that the false detection rate in S2XLS XMM-LSS is higher than expected at a detection threshold of $5\sigma$ \citep[comparing to previous SCUBA-2 surveys, e.g.,][]{Geach2017,Simpson2019}. This is possibly motivated by the steep slope of the $850\mu{\rm m}$ number counts; at $>15\,{\rm mJy}$ there are very few real sources and so even a small number of spurious detections dramatically increases the false detection rate.     
 
\section{DISCUSSION}
\subsection{Bright-end number counts of the $\mathbf{ 850}\,\mu\rm{\mathbf m}$ population}\label{section:number counts}
In Table \ref{tab:catalogue} we present the S2LXS XMM-LSS catalogue, which contains $40$ sources with intrinsic flux densities in the range $7$ to $48\,{\rm mJy}$ detected at a significance of ${>}5\sigma$.  This catalogue includes observed and deboosted flux densities, completeness levels, positional uncertainties and purity rates.

 \begin{table*}
 	\caption{S2LXS XMM-LSS catalogue.  This catalogue contains $40$ sources detected at a significance of ${>}5\sigma$. Coordinates are J2000. The $S^{\rm obs}_{\rm 850\mu m}\pm \sigma_{\rm inst}$ column gives the observed flux density and instrumental noise, the S/N column gives the signal-to-noise ratio of the detection, and the $S^{\rm deb}_{\rm 850\mu m}\pm \sigma_{\rm tot}$ column gives the estimated deboosted (true) flux density and combined total (instrumental and deboosting) uncertainty. Column $C$ gives the completeness level, $\Delta\theta$ the positional uncertainty (calculated from equation \ref{eq: pos_uncert}) and the final column $P$ gives the purity rate.}
 	\begin{threeparttable}
	\label{tab:catalogue}
	\begin{tabular}{>{\centering\arraybackslash}p{3.4cm}>{\centering\arraybackslash}p{1.4cm}>{\centering\arraybackslash}p{1.4cm}>{\centering\arraybackslash}p{1.6cm}>{\centering\arraybackslash}p{1.1cm}>{\centering\arraybackslash}p{1.6cm}>{\centering\arraybackslash}p{1.1cm}>{\centering\arraybackslash}p{1.1cm}>{\centering\arraybackslash}p{1.1cm}} 
	\hline \\ [-4pt]
		 S2LXS ID & RA & Dec & $S^{\rm obs}_{\rm 850\mu m}\pm \sigma_{\rm inst}$ & $S/N$ & $S^{\rm deb}_{\rm 850\mu m}\pm \sigma_{\rm tot}$ & $C$ & $\Delta\theta$&$P$ \\[4pt]
		 &  &  & [mJy] & & [mJy] & & arcsec & \\[4pt]
		\hline \\ [-5pt]
\,\,\,S2LXSJ021831-053131$^{ad}$&02 18 30.7&-05 31 31.3&$49.7\pm{4.3}$&$11.5$&$48.1\pm{6.3}$&$1.00$&$0.80$&$1.00$ \\[1pt]
S2LXSJ022700-052405&02 27 00.3&-05 24 05.2&$26.0\pm{4.5}$&$5.7$&$20.3\pm{7.4}$&$0.37$&$1.85$&$1.00$ \\[1pt]
S2LXSJ021735-054854&02 17 34.9&-05 48 53.6&$25.9\pm{4.8}$&$5.4$&$18.5\pm{7.9}$&$0.19$&$1.99$&$0.97$ \\[1pt]
\,\,\,S2LXSJ022548-041751$^{b}$&02 25 47.7&-04 17 51.5&$25.9\pm{3.3}$&$7.7$&$23.1\pm{5.4}$&$0.97$&$1.29$&$1.00$ \\[1pt]
S2LXSJ021520-053222&02 15 19.9&-05 32 21.8&$25.4\pm{4.6}$&$5.5$&$18.4\pm{7.7}$&$0.22$&$1.95$&$0.99$ \\[1pt]
S2LXSJ022252-042412&02 22 51.8&-04 24 12.0&$25.3\pm{4.9}$&$5.2$&$16.8\pm{8.0}$&$0.10$&$2.09$&$0.93$ \\[1pt]
S2LXSJ022324-042602&02 23 23.8&-04 26 01.7&$25.1\pm{4.7}$&$5.4$&$17.9\pm{7.7}$&$0.19$&$1.99$&$0.98$ \\[1pt]
S2LXSJ022735-051558&02 27 35.2&-05 15 58.0&$23.8\pm{4.1}$&$5.9$&$18.6\pm{6.5}$&$0.43$&$1.79$&$1.00$ \\[1pt]
S2LXSJ022322-033747&02 23 22.1&-03 37 47.5&$21.5\pm{4.1}$&$5.3$&$14.1\pm{6.8}$&$0.10$&$2.04$&$0.95$ \\[1pt]
S2LXSJ022614-045616&02 26 13.7&-04 56 15.5&$21.0\pm{3.7}$&$5.6$&$15.2\pm{6.2}$&$0.25$&$1.89$&$1.00$ \\[1pt]
S2LXSJ022854-042445&02 28 54.0&-04 24 45.2&$20.0\pm{3.6}$&$5.6$&$13.8\pm{5.9}$&$0.19$&$1.90$&$1.00$ \\[1pt]
S2LXSJ021826-041900&02 18 26.4&-04 18 59.1&$19.9\pm{3.8}$&$5.2$&$13.1\pm{6.2}$&$0.10$&$2.07$&$0.94$ \\[1pt]
S2LXSJ022342-044501&02 23 41.8&-04 45 01.0&$19.8\pm{3.5}$&$5.7$&$14.0\pm{5.9}$&$0.23$&$1.85$&$1.00$ \\[1pt]
S2LXSJ022332-042948&02 23 31.5&-04 29 47.6&$19.8\pm{3.9}$&$5.1$&$12.7\pm{6.3}$&$0.08$&$2.11$&$0.92$ \\[1pt]
S2LXSJ022649-035146&02 26 49.3&-03 51 46.0&$19.7\pm{3.4}$&$5.8$&$14.1\pm{5.8}$&$0.26$&$1.83$&$1.00$ \\[1pt]
S2LXSJ021938-042930&02 19 37.9&-04 29 30.4&$19.4\pm{3.1}$&$6.3$&$14.9\pm{5.3}$&$0.47$&$1.65$&$1.00$ \\[1pt]
S2LXSJ022400-034623&02 24 00.3&-03 46 22.6&$19.1\pm{3.4}$&$5.6$&$13.4\pm{5.8}$&$0.21$&$1.89$&$1.00$ \\[1pt]
S2LXSJ022042-053806&02 20 42.0&-05 38 05.7&$18.6\pm{2.4}$&$7.9$&$16.4\pm{4.1}$&$0.96$&$1.26$&$1.00$ \\[1pt]
\,\,\,S2LXSJ021943-052436$^{c}$&02 19 42.8&-05 24 35.8&$18.4\pm{2.0}$&$9.3$&$16.7\pm{3.7}$&$0.99$&$1.03$&$1.00$ \\[1pt]
S2LXSJ022256-051816&02 22 56.1&-05 18 15.6&$17.9\pm{3.6}$&$5.0$&$11.5\pm{5.5}$&$0.07$&$2.16$&$0.90$ \\[1pt]
S2LXSJ022044-044641&02 20 43.8&-04 46 40.8&$17.8\pm{3.0}$&$6.0$&$12.8\pm{5.2}$&$0.28$&$1.74$&$1.00$ \\[1pt]
S2LXSJ022508-045914&02 25 07.5&-04 59 13.7&$17.4\pm{3.5}$&$5.0$&$10.6\pm{5.3}$&$0.05$&$2.16$&$0.91$ \\[1pt]
S2LXSJ022122-042315&02 21 21.9&-04 23 15.1&$17.0\pm{3.2}$&$5.4$&$11.0\pm{5.1}$&$0.11$&$1.98$&$0.98$ \\[1pt]
S2LXSJ022426-042736&02 24 26.2&-04 27 36.1&$17.0\pm{3.3}$&$5.2$&$10.8\pm{5.1}$&$0.08$&$2.07$&$0.94$ \\[1pt]
S2LXSJ021501-050817&02 15 00.6&-05 08 16.6&$16.8\pm{3.2}$&$5.3$&$10.6\pm{5.1}$&$0.08$&$2.04$&$0.95$ \\[1pt]
S2LXSJ021536-045218&02 15 36.4&-04 52 17.9&$15.9\pm{3.1}$&$5.2$&$10.0\pm{4.8}$&$0.07$&$2.08$&$0.94$ \\[1pt]
\,\,\,S2LXSJ021921-045651$^{d}$&02 19 20.9&-04 56 51.1&$14.2\pm{2.3}$&$6.1$&$10.3\pm{3.9}$&$0.36$&$1.73$&$1.00$ \\[1pt]
S2LXSJ022039-053937&02 20 39.3&-05 39 37.3&$14.2\pm{2.6}$&$5.5$&$9.5\pm{4.1}$&$0.16$&$1.94$&$1.00$ \\[1pt]
S2LXSJ021957-053406&02 19 56.5&-05 34 06.4&$13.7\pm{2.5}$&$5.6$&$9.4\pm{4.0}$&$0.19$&$1.91$&$1.00$ \\[1pt]
S2LXSJ022032-053615&02 20 31.7&-05 36 15.3&$13.6\pm{2.3}$&$5.8$&$9.6\pm{3.7}$&$0.28$&$1.80$&$1.00$ \\[1pt]
S2LXSJ021938-052502&02 19 37.9&-05 25 02.1&$13.4\pm{2.1}$&$6.5$&$9.9\pm{3.5}$&$0.46$&$1.60$&$1.00$ \\[1pt]
S2LXSJ022134-053558&02 21 34.5&-05 35 58.2&$13.3\pm{2.3}$&$5.7$&$9.3\pm{3.7}$&$0.26$&$1.84$&$1.00$ \\[1pt]
\,\,\,S2LXSJ021907-052202$^{d}$&02 19 07.1&-05 22 01.6&$12.7\pm{2.5}$&$5.1$&$8.5\pm{3.9}$&$0.10$&$2.14$&$0.91$ \\[1pt]
\,\,\,S2LXSJ021910-051532$^{d}$&02 19 09.8&-05 15 32.0&$12.4\pm{2.4}$&$5.1$&$8.4\pm{3.7}$&$0.11$&$2.12$&$0.92$ \\[1pt]
\,\,\,S2LXSJ021939-052315$^{d}$&02 19 39.3&-05 23 15.0&$11.1\pm{2.0}$&$5.6$&$7.5\pm{3.1}$&$0.15$&$1.89$&$1.00$ \\[1pt]
S2LXSJ022122-051113&02 21 22.4&-05 11 13.0&$11.0\pm{2.2}$&$5.1$&$7.6\pm{3.3}$&$0.12$&$2.14$&$0.91$ \\[1pt]
S2LXSJ022018-051124&02 20 18.4&-05 11 23.9&$10.9\pm{1.8}$&$6.1$&$7.7\pm{3.0}$&$0.18$&$1.72$&$1.00$ \\[1pt]
S2LXSJ022150-052835&02 21 49.8&-05 28 34.6&$10.8\pm{2.1}$&$5.3$&$7.4\pm{3.1}$&$0.13$&$2.03$&$0.96$ \\[1pt]
S2LXSJ022133-051558&02 21 33.2&-05 15 57.5&$10.4\pm{2.0}$&$5.2$&$7.2\pm{3.1}$&$0.12$&$2.09$&$0.93$ \\[1pt]
S2LXSJ022003-052000&02 20 03.1&-05 19 59.5&$9.8\pm{1.8}$&$5.6$&$6.8\pm{2.9}$&$0.09$&$1.91$&$1.00$ \\[1pt]

\hline
\end{tabular}
\begin{tablenotes}
\item[$a$] Ultra-bright lensed submillimetre galaxy known as `Orochi' \citep{Ikarashi2011}.
\item[$b$] Lensed hyper-luminous infrared galaxy \citep[HXMM05; ][]{Oliver2012,Bussmann2015}. 
\item[$c$] The SCUBA-2 source has five ALMA counterparts \citep{Oliver2012,Bussmann2015}.
\item[$d$] S2CLS UKIDSS-UDS source \citep{Geach2017} with ALMA follow up observations \citep{Stach2019}.
\end{tablenotes}
\end{threeparttable}
\end{table*}

Number counts describe the surface density of sources per observed flux density interval ($dN/dS$), providing a simple measure of source abundance and a valuable tool to compare with model predictions.  We estimate the $850\,\mu{\rm m}$ number counts for S2LXS XMM-LSS from the survey catalogue, correcting for the effects of flux boosting, incompleteness and the probability that an individual source is a false detection.  We provide a brief description of this process here.  For each of the $40$ sources in the ${>}5\sigma$ catalogue (Table \ref{tab:catalogue}) we first estimate the deboosted (true) flux density by drawing a random sample from a \textit{p}($S_{\rm true}$) distribution (estimated from our simulated maps, see Section \ref{section: flux boosting} for details) consistent with the observed flux density and local instrumental noise of the source. We then apply a completeness correction, retrieving the completeness rate (which is estimated in bins of deboosted flux density and local instrumental noise) via $2$-dimensional spline interpolation from the look up table generated in section \ref{section:completeness}.  Next we correct for the probability that a source is spurious based on the signal-to-noise ratio of the source detection (see Section \ref{section: false_detection}).  Once we have applied these corrections we evaluate the corrected counts for the S2LXS XMM-LSS catalogue in flux intervals of $\Delta{\rm log}_{\rm 10}(S) = 0.1$, and then divide by the area of the S2LXS XMM-LSS science map\footnote{To calculate the number counts per square degree we divide by the full area of the S2LXS XMM-LSS mosaic, as given the statistical nature of flux boosting a source in even the lowest flux bin could be found at any position in the S2LXS XMM-LSS mosaic.} to measure the number counts ($dN/dS$) per flux interval per square degree.  Since the observed flux density of a source maps onto a range of intrinsic flux densities, we evaluate $dN/dS$ $1000$ times (following the process above), each time randomly drawing the deboosted flux density for a source from the full intrinsic flux distribution (\textit{p}$(S_{\rm true})$).  We take the final number counts as the mean of $dN/dS$ across the $1000$ realisations and the standard deviation of $dN/dS$ in each flux bin as an additional uncertainty \citep[to the Poisson error;][]{Gehrels1986}.  We note that the corrections for flux boosting and completeness are estimated in bins of flux density and instrumental noise to account for the variation in the instrumental noise in the S2LXS XMM-LSS mosaic.

The purpose of the various corrections we make is to recover the `true' underlying source distribution. However, the simulations we use to derive the corrections are imperfect; i.e., the source count model we use at $S_{850\,\mu{\rm m}}>15\,{\rm mJy}$ is based on the observed source counts of the S2LXS XMM-LSS survey. Therefore, it is important to test if any systematic biases remain.  We also note that owing to the variation in instrumental noise across the S2LXS XMM-LSS mosaic we detect $95\%$ of sources in deeper regions of the map where at least $2$ PONG tiles overlap, and so it is also crucial to test whether the location of sources detected in our map will influence our results. 

To do this we inject a realistic source count model into our jackknife map (see section \ref{section:simulated maps} for details of how the jackknife map is created). The faint-end of the number counts is well constrained and so we use the \citet{Schechter1976} fit from \citet{Geach2017} to model the source counts below $15\,{\rm mJy}$. To produce a realistic number counts model above $15\,{\rm mJy}$ (noting that the bright-end of the $850\,\mu{\rm m}$ number counts is not well constrained in previous work) we fit a simple power law of the form $y=ax^k$ to the S2LXS XMM-LSS corrected number counts (see equation \ref{eq: num_fit}). We make an arbitary distinction between the `faint' and `bright' source count models at $15\,{\rm mJy}$ as this is consistent with the S2CLS results, which show an upturn in the $850\,\mu{\rm m}$ number counts above this flux density \citep{Geach2017}.  We recover sources in an identical manner to the real data, adopting the formal detection threshold of the S2LXS XMM-LSS survey (${>}5\sigma$). We then apply the various corrections (Sections \ref{section: flux boosting}, \ref{section:completeness} and \ref{section: false_detection}) to estimate the number counts. By comparing the recovered counts (before and after corrections) to the exact model counts injected into the maps we can test the effectiveness of our corrections. To check that our number counts corrections are not overly biased by the input source count model, we repeat the process above for two more extreme source count models (at the bright-end) by varying the slope $k$ of the power law fit (in equation \ref{eq: num_fit}), using a model with a steeper slope $k=-6.60$ and a model with a shallower slope $k=-4.60$.  We produce $100$ simulated maps for each source count model, and take an average of the counts before and after the corrections. 

\begin{figure*}
	\includegraphics[width=1\linewidth]{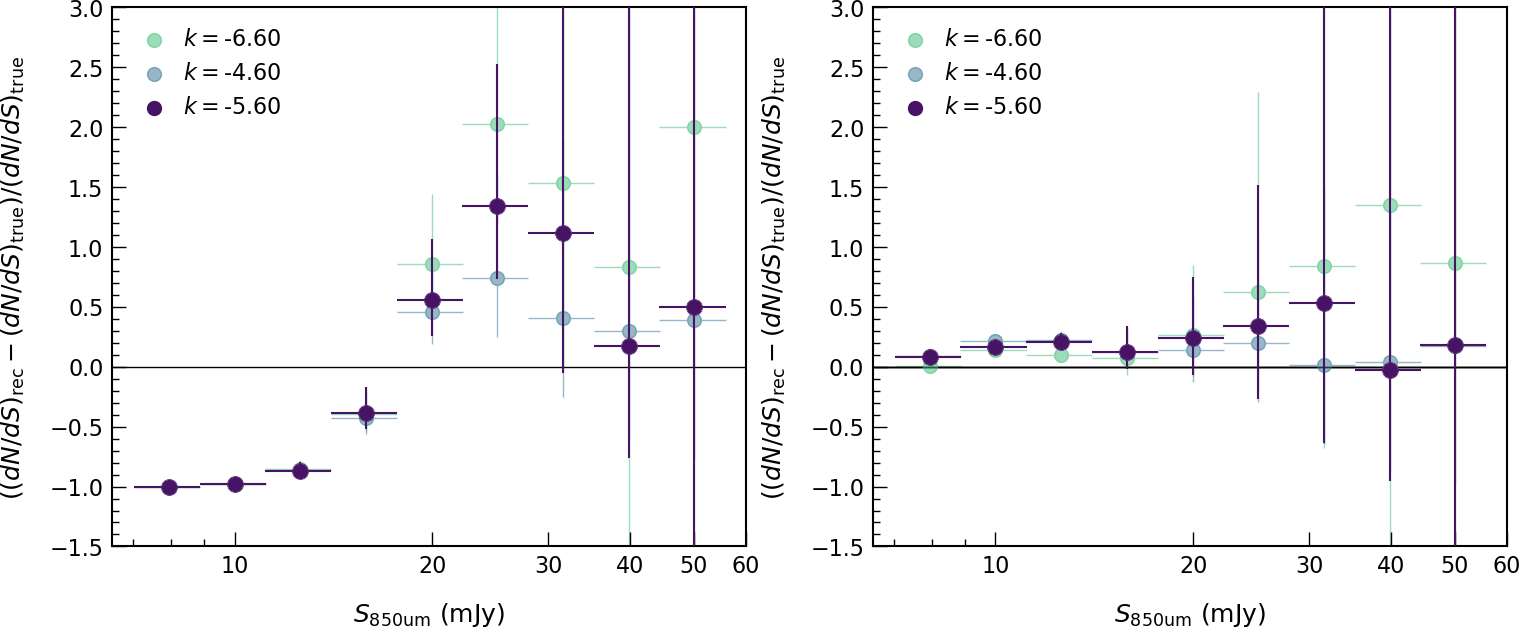}
    \caption{Comparison of recovered number counts to the `true' number counts in simulated maps. We inject a realistic source count model into a jackknife realisation of the S2LXS XMM-LSS mosaic and then recover sources detected at ${>5}\sigma$ (the formal detection threshold of S2LXS XMM-LSS).  To check that our number counts corrections are not overly biased by the input source count model we repeat this for two more extreme models by varying the slope $k$ of the power law that we use to model the bright-end source count. In the left panel the recovered counts are not corrected for the effects of flux boosting, incompleteness or the probability that a source is a false detection.  We show that in the absence of corrections the number counts are under-estimated by up to $99$ per cent in the faintest flux density bins ($S_{\rm 850\mu m}\leq18\,{\rm mJy}$) in which incompleteness dominates. At brighter flux densities the number counts are generally over-estimated due to flux boosting. In the right panel we show these counts with the various corrections applied, illustrating that the `true' number counts can be robustly recovered.  We note that in the brightest flux density bins ($S_{\rm 850\mu m}>30\,{\rm mJy}$) the corrections are less effective, likely due to the low number statistics in these bins (evidenced by the large Poisson errors). The error bars in both panels show Poisson uncertainties \citep{Gehrels1986} only.}
    \label{fig:number_counts_diff}
\end{figure*}

In Figure \ref{fig:number_counts_diff} we show the average $((dN/dS)_{\rm rec}-(dN/dS)_{\rm true})/(dN/dS)_{\rm true}$ before and after corrections have been applied for all three source count models.  In the absence of corrections the number counts are under-estimated by up to $99$ per cent in the faintest flux bins ($S_{\rm 850\mu m}\leq18\,{\rm mJy}$), largely influenced by the incompleteness of the S2LXS XMM-LSS survey at these flux densities. For brighter flux bins we generally see an over-estimation of the number counts due to flux boosting. Once we apply corrections the right-hand panel of Figure \ref{fig:number_counts_diff} shows that we can reliably recover the true source counts for all three source count models, noting that for the brightest flux bins  $S_{\rm 850\mu m}>30\,{\rm mJy}$ the corrections appear less effective, likely due to low number statistics (evidenced by the large Poisson errors).  

\begin{figure*} 
	\includegraphics[width=1\linewidth]{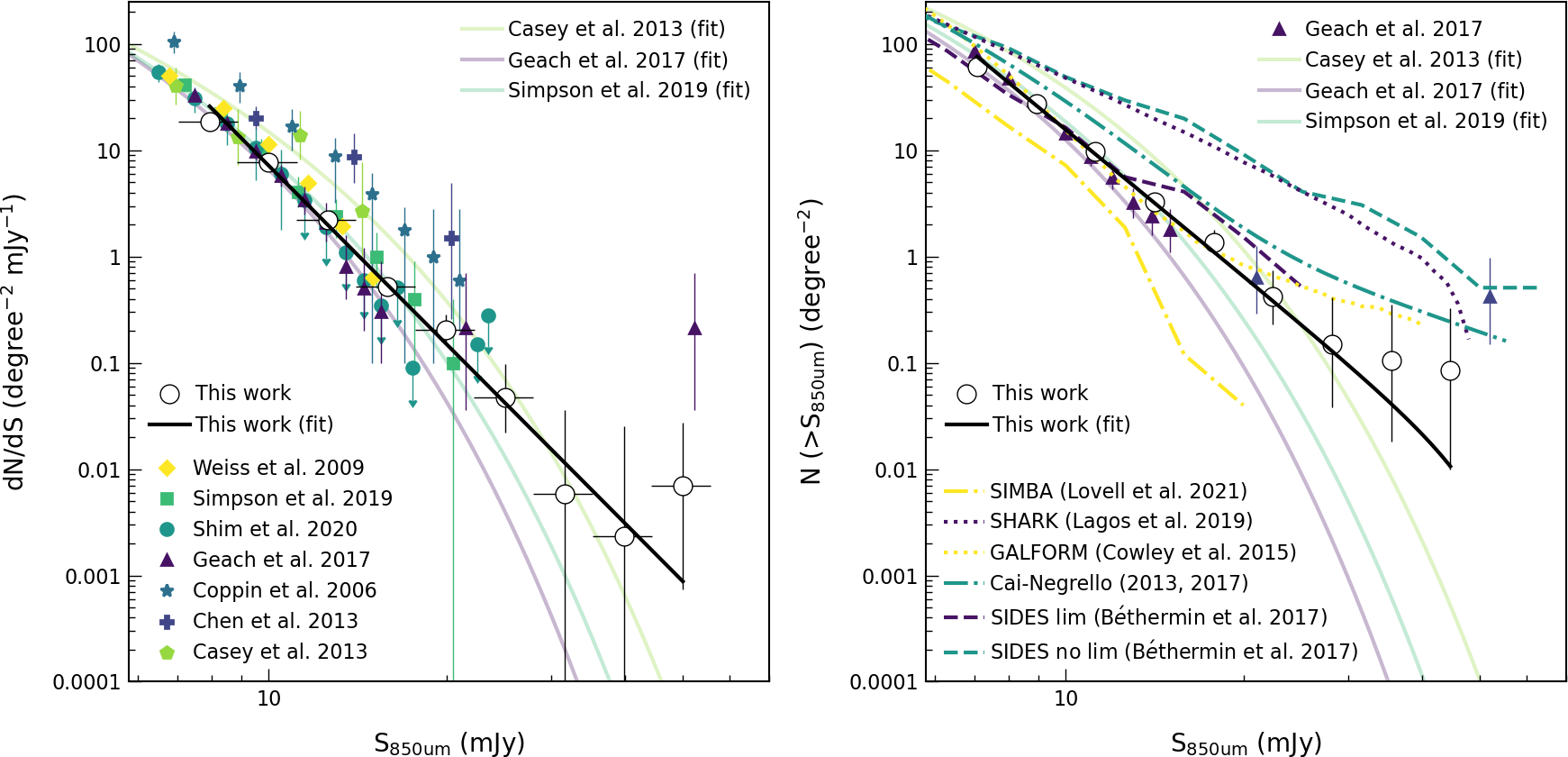}
    \caption{Number counts of $850\,\mu{\rm m}$ sources measured from the S2LXS XMM-LSS survey $5\sigma$ catalogue (white circles with black outline). In the left panel we present the differential number counts from this work along with observational constraints from S2CLS \citep{Geach2017}, S2COSMOS \citep{Simpson2019} and the other main blank-field surveys \citep{Casey2013,Coppin2006,shim2020,Weiss2009}, as well as from \citet{Chen2013} who combine data from two cluster lensing fields and three blank fields. The S2XLS XMM-LSS differential number counts are consistent with these previous studies within the uncertainties for the flux range probed. We find an excess in the number counts (above the \citet{Schechter1976} function fit of \citet{Geach2017}) at ${\gtrsim}15\,{\rm mJy}$, similar to that found in S2CLS \citep{Geach2017}. In the right panel we present the S2LXS XMM-LSS cumulative number counts, comparing both to observations and several simulated galaxy models. For clarity we only plot the observational constraints from S2LXS XMM-LSS and S2CLS, and the model fits from S2COSMOS and \citet{Casey2013}. We show the $850\,\mu{\rm m}$ number counts from semi-analytic models \citep{cowley2015,lagos2019}, empirical models \citep{bethermin2017,Cai2013,negrello2017} and the cosmological hydrodynamical simulation {\sc Simba} \citep{Lovell2021}. The results from \citet{Lovell2021} and \citet{cowley2015} both take into account blending from a $15$\,arcsec beam. The \citet{bethermin2017} standard model has a star-formation rate limit of $1000\,{\rm M}_\odot\,{\rm yr}^{-1}$. We show the results both with and without this limit. The \citet{bethermin2017} models shown do not take account of blending, but do include a prescription for lensing. The Cai-Negrello model combines the counts of unlensed and strongly-lensed galaxies. The excess in the S2LXS XMM-LSS number counts at ${\gtrsim}15\,{\rm mJy}$ is most consistent with the results from the Cai-Negrello empirical model \citep{Cai2013,negrello2017}, the semi-analytic model of \citet{cowley2015}, and the standard empirical model of \citet{bethermin2017}. The error bars in both panels show Poisson uncertainties \citep{Gehrels1986}.}
    \label{fig:number_counts}
\end{figure*}

The S2LXS XMM-LSS $850\,\mu{\rm m}$ differential and cumulative number counts are presented in Table \ref{tab:number counts} and Figure \ref{fig:number_counts}.  We show our results for intrinsic flux densities ${>}7\,{\rm mJy}$, with the minimum flux consistent with the deboosted flux density of the faintest source in the S2LXS XMM-LSS catalogue.  Since the bright-end (${>}15\,{\rm mJy}$) of the $850\,\mu{\rm m}$ differential number counts are no longer well represented by a \citet{Schechter1976} function we fit the S2LXS XMM-LSS $850\,\mu{\rm m}$ differential number counts (see Table \ref{tab:number counts} for values) with a power law (Figure \ref{fig:number_counts}, solid black line):

 \begin{ceqn}
\begin{align}
\frac{dN}{dS} = (2.8 \pm{0.3}) \times 10^6  \times \left(\frac{{S_{\rm 850\mu m}}}{{\rm mJy}}\right)^{-5.6 \pm 0.5}. 
\label{eq: num_fit}
\end{align}
\end{ceqn}

 In Figure \ref{fig:number_counts} we also plot the observational constraints from S2CLS \citep{Geach2017}, S2COSMOS \citep{Simpson2019} and the other main $850\mu{\rm m}$ wide-area surveys \citep[e.g.,][]{Casey2013,Chen2013,Coppin2006,shim2020,Weiss2009} for comparison. The results from S2XLS XMM-LSS are in good agreement with these previous surveys for the flux range probed.   We note that since the S2LXS XMM-LSS survey has a contiguous survey area of $9\,{\rm deg}^2$ cosmic variance will have a smaller affect on our number count estimates compared to previous work, with \citet{Simpson2019} finding no significant variation in the $850\,\mu{\rm m}$ number counts on scales of $0.5$--$3\,{\rm deg}^2$. 
 
 Owing to a slight overlap in the S2LXS XMM-LSS and S2CLS UKIDSS-UDS maps these surveys have $5$ sources in common.  One of these sources is an ultra-bright lensed galaxy \citep[`Orochi';][]{Ikarashi2011}; with a flux density of $S_{\rm 850\mu m}\simeq50\,{\rm mJy}$ this is the brightest extra-galactic source detected in either survey. \citet{Geach2017} find an excess in the S2CLS differential number counts (i.e., above the \citet{Schechter1976} function fit) at $S_{\rm 850\mu m}>15\,{\rm mJy}$ (see Figure \ref{fig:number_counts}). We find a similar upturn in the S2LXS XMM-LSS differential number counts, but with a survey area of $9\,{\rm deg}^2$ we detect double the number of $S_{\rm 850\mu m}>15\,{\rm mJy}$ extragalactic sources (with $11$ sources detected at ${>}5\sigma$), and so we are able to reduce the Poisson errors on these measurements.  The S2LXS XMM-LSS catalogue only includes one source with an observed flux density $S_{\rm 850\mu m}>30\,{\rm mJy}$, and so our survey is hampered by low number statistics in this flux regime.  This,  coupled with our correction for the effects of flux boosting\footnote{We evaluate $dN/dS$ $1000$ times each time randomly drawing the deboosted flux density for a source from a \textit{p}($S_{\rm true}$) distribution (estimated from our simulated maps, see Section \ref{section: flux boosting} for details) consistent with the observed flux density and local instrumental noise of the source.  As we are sampling the full intrinsic flux distribution for a source, over the course of $1000$ iterations, the estimated deboosted flux densities may map to more than one flux density bin.  This is the reason we recover a number counts estimate for intrinsic flux densities $30\,{\rm mJy}<S_{850\,\mu{\rm m}}<48\,{\rm mJy}$ despite no sources with observed flux densities in this range.}, means we see large Poisson uncertainties on our number counts estimates at flux densities above $30\,{\rm mJy}$. 
 
 An excess at the bright end of the number counts is also observed in studies at shorter \citep[e.g., \textit{Herschel $500\,\mu{\rm m}$};][]{negrello2010,wardlow2013} and longer  \citep[e.g., South Pole Telescope $1.4\,{\rm mm}$; ][]{Vieira2010,Mocanu2013} wavelengths, and is attributed to the presence of local objects and high-redshift gravitationally lensed sources \citep[e.g.,][]{negrello2010,Vieira2010}.  S2CLS provides tentative evidence that the over-abundance of sources at the bright-end of the $850\mu{\rm m}$ number counts is similarly influenced \citep{Geach2017}; of the three sources detected above $20\,{\rm mJy}$ in S2CLS the brightest ($S_{\rm 850\mu m}\simeq200\,{\rm mJy}$) is a well-known Galactic object in the \textit{Akari}-North Ecliptic Pole (the Cat's Eye Nebula) and another is  a known ultra-bright lensed submillimetre galaxy \citep[`Orochi'][]{Ikarashi2011} which lies at $z_{\rm phot}\simeq3.5$, \citep{Dudzevi19}. The S2XLS XMM-LSS ${>}5\sigma$ catalogue includes at least two gravitationally lensed sources; `Orochi' \citep{Ikarashi2011} and a high-redshift ($z_{\rm spec}\sim3$) hyper-luminous infrared galaxy from HerMES \citep{Oliver2012,Bussmann2015,Leung2019}.  We visually inspect the available multi-wavelength imaging (see Table \ref{tab:multi-wavelength data}) centred on the position of each S2LXS XMM-LSS source and query the NASA/IPAC Extragalactic Database (NED) for nearby $z<0.1$ sources, allowing us to confirm that there are no Galactic objects in the S2LXS XMM-LSS catalogue.  We also search the multi-wavelength imaging and NED around the position of each S2LXS XMM-LSS source to identify bright, massive foreground galaxies, which are indicative of strong lensing.  We find that for $16$ ($40$ per cent) sources in the S2LXS XMM-LSS catalogue there is a bright, massive foreground galaxy within the SCUBA-2 beam.  Whilst it is reasonable to expect (based on previous work) that the over-abundance of sources observed at high flux densities in the S2LXS XMM-LSS survey is, at least in part, motivated by the presence high-redshift lensed galaxies, we cannot rule out a contribution from intrinsically bright sources \citep[i.e., similar to GN20 a $S_{850\,\mu{\rm m}}>20\,{\rm mJy}$ galaxy detected in the SCUBA survey of the GOODS-North field;][]{Pope2005,daddi2009} or proto-cluster core type objects such as the Distant Red Core \citep[DRC;][]{ivison2016,oteo2018}.  It is also well known that a fraction of single dish $850\,\mu{\rm m}$ selected sources are the product of multiple fainter submillimetre galaxies that are blended in the coarse $15$\,arcsec resolution of the SCUBA-2 beam \citep[e.g.,][]{Simpson2015,Stach2018}.  The multiplicity fraction increases with the brightness of the single dish source, with an estimated multiplicity fraction of $44$ per cent for single dish $850\,\mu{\rm m}$ selected sources with flux densities $S_{\rm 850\mu m}>9\,{\rm mJy}$ \citep[e.g.,][]{Stach2018}.  Surveys following up observations of single dish selected $850\mu{\rm m}$ sources with sub-arcsecond resolution interferometers find a systematic reduction in the estimated number counts by a factor of approximately $41$ per cent at $S_{\rm 870\mu m}\geq7\,{\rm mJy}$ and $24$--$30$ per cent at $S_{\rm 870\mu m}\geq12\,{\rm mJy}$, though  due to small number statistics the latter is not well constrained \citep[e.g.,][]{Hill2018,Stach2018}.  We will explore the nature of the S2LXS XMM-LSS sources via follow up ALMA observations (Project ID: 2022.1.01030.S, PI: T. K. Garratt) of the brightest sources in the S2LXS XMM-LSS catalogue.  With the exquisite resolution of ALMA we expect to pinpoint the position of multi-wavelength counterparts to the brightest S2LXS XMM-LSS sources, allowing us to constrain the relative contributions of source blending, lensing and intrinsically bright sources to the bright-end of the $850\,\mu{\rm m}$ number counts.

 \begin{table*}
	\centering
	\caption{Differential ($dN/dS$) and cumulative ($N>S'$) $850\,\mu{\rm m}$ number counts measured from the S2LXS XMM-LSS survey.  We evaluate $dN/dS$ $1000$ times taking the mean as the final number counts estimate.  The flux density bin widths are $\Delta{\rm log}_{10}(S)=0.1$, with $S$ and $S'$  the flux bin centres and edges respectively.   The first set of errors shown are the Poisson uncertainties \citep{Gehrels1986} and the second set are the standard deviation on the number counts estimates over the $1000$ realisations (see text for details).}
	\label{tab:number counts}
	\begin{tabular}{>{\centering\arraybackslash}p{1.80cm}>{\centering\arraybackslash}p{1.15cm}>{\centering\arraybackslash}p{1.15cm}>{\centering\arraybackslash}p{3.1cm}>{\centering\arraybackslash}p{2.9cm}} 
	\hline \\ [-4pt]
		${\rm log}_{\rm 10}(S/{\rm mJy}) $ & $S$ &  $S'$ &  $dN/dS$ &  $N>S'$\\[1pt]
		& [mJy] & [mJy] & [$10^{-3}$\,deg$^{-2}$\,mJy$^{-1}$] & [$10^{-3}$\,deg$^{-2}$]\\[4pt]
		\hline \\ [-4pt]
$0.9$ & $7.9$ & $7.1$ & $18,637\substack{+203 \\ -131}\pm28,569$ & $61,807\substack{+735 \\ -618}\pm53,964$ \\[4pt] 
$1.0$ & $10.0$ & $8.9$ & $7,762\substack{+179 \\ -123}\pm6,564$ & $27,644\substack{+684 \\ -566}\pm14,965$ \\[4pt]  
$1.1$ & $12.6$ & $11.2$ & $2,221\substack{+138 \\ -93}\pm1,460$ & $9,733\substack{+605 \\ -485}\pm4,056$ \\ [4pt] 
$1.2$ & $15.8$ & $14.1$ & $525\substack{+103 \\ -67}\pm291$ & $3,281\substack{+520 \\ -398}\pm981$ \\[4pt]  
$1.3$ & $20.0$ & $17.8$ & $204\substack{+79 \\ -50}\pm91$ & $1,361\substack{+433 \\ -306}\pm403$ \\[4pt]  
$1.4$ & $25.1$ & $22.4$ & $47\substack{+50 \\ -25}\pm28$ & $421\substack{+329 \\ -191}\pm165$ \\ [4pt] 
$1.5$ & $31.6$ & $28.2$ & $6\substack{+30 \\ -6}\pm9$ & $149\substack{+264 \\ -111}\pm64$ \\ [4pt] 
$1.6$ & $39.8$ & $35.5$ & $2\substack{+23 \\ -2}\pm5$ & $106\substack{+248 \\ -88}\pm13$ \\ [4pt] 
$1.7$ & $50.1$ & $44.7$ & $7\substack{+21 \\ -6}\pm4$ & $85\substack{+240 \\ -75}\pm44$ \\ [4pt]

		\hline
	\end{tabular}
\end{table*}
\subsection{Comparison to models}

In Figure \ref{fig:number_counts} we compare our results to: semi-analytic models {\sc Galform} \citep{cowley2015,lacey2016} and {\sc Shark} \citep{lagos2019}; empirical models {\sc Sides} \citep{bethermin2017} and the Cai-Negrello model \citep{Cai2013,negrello2017}; and the cosmological hydrodynamical simulation {\sc Simba} \citep{Lovell2021}. These models are all able to broadly reproduce the faint-end of the $850\,\mu{\rm m}$ number counts \citep[e.g.,][]{Geach2017,Simpson2019}.  

The blending of multiple faint submillimetre galaxies in the coarse $15$\,arcsec resolution of the SCUBA-2 beam is known to boost estimates of the $850\,\mu{\rm m}$ number counts, with follow up surveys (using sub-arcsecond resolution interferometers) finding a systematic reduction in the estimated number counts by a factor of approximately $41$ per cent at $S_{\rm 870\mu m}\geq7\,{\rm mJy}$.  Of the models discussed here the semi-analysic model {\sc Galform}, the empirical model {\sc Sides} and the hydrodynamical simulation {\sc Simba} all simulate the effects of source blending in a $15$\,arcsec beam, which allows us to make a more `like-for-like' comparison with the observed S2LXS XMM-LSS number counts.  \citet{Lovell2021} use a 3D dust continuum radiative transfer code to estimate the submillimetre fluxes of galaxies in the cosmological hydrodynamic simulation {\sc Simba}. To model the effects of source blending a projected {\sc Simba} $850\,\mu{\rm m}$ lightcone is convolved with the SCUBA-2 point spread function \citep{Dempsey2013} to produce an `observed' SCUBA-2 map. The {\sc Simba} predicted $850\,\mu{\rm m}$ number counts are more than a factor of $2$ below the observed S2LXS XMM-LSS counts.  The semi-analytic model {\sc Galform} \citep{lacey2016} invokes a mildly top-heavy initial mass function in starbursts to reproduce the observed number counts and redshift distribution \citep[e.g.,][]{simpson2014,wardlow2011} of $850\,\mu{\rm m}$ selected sources \citep[although we note this model is unable to reproduce the redshift distribution of bright SMGs with flux densities $S_{850\,\mu{\rm m}}>9{\,{\rm mJy}}$;][]{Chen2022}. Sources are extracted from the simulated maps using a top-down peak-finding approach and the number counts derived from the source-extracted catalogue. The {\sc Galform} model is in good agreement with the observed S2XLS XMM-LSS counts within the uncertainties.

The upturn in the 850um number counts is motivated, at least in part, by the presence of high-redshift gravitationally lensed sources.  A prescription for lensing is included in the empirical model of  \citet{bethermin2017}, which uses an updated version of the two star-formation modes \citep{bethermin2012,sargent2012} galaxy evolution model to derive global galaxy properties using empirical relations from observations. Taking the galaxy properties generated from this model, \citet{bethermin2017} use an abundance matching technique to populate dark-matter halos with galaxies to produce a $2\,{\rm deg}^2$ simulation of the extragalactic sky. The simulation includes a simple estimate of lensing that depends on redshift but does not account for the mass distributions in the foreground. Intrinsic number counts are estimated from the simulated catalogue. The standard model includes a sharp star-formation limit of $1000\,{\rm M}_\odot\,{\rm yr}^{-1}$. In Figure \ref{fig:number_counts} we plot the predicted intrinsic counts from \cite{bethermin2017} for the standard model and the model with no star-formation limit. We note that \citet{bethermin2017} also estimate the number counts from a source-extracted catalogue to take into account source blending. The source-extracted counts mirror the intrinsic counts at bright flux densities and so for clarity we do not plot the source extracted counts in Figure \ref{fig:number_counts}. The standard model is broadly in agreement with the S2LXS observed number counts below $30\,{\rm mJy}$, but fails to reproduce the observed counts at higher flux densities. The model without a star-formation limit significantly over-predicts the counts by up to a factor $20$ across the flux density range probed, which indicates that including a star-formation limit in some form is necessary for this model to reproduce the observed counts. 
The Cai-Negrello empirical model \citep{Cai2013,negrello2017} also takes into account the influence of strongly lensed galaxy, by combining theoretical counts for late-type, unlensed and strongly lensed galaxies. The maximum magnification assumed for calculating the strongly lensed number counts is $\mu_{\rm max}=15$. We note that the Cai model \citep{Cai2013} presented here has been re-calibrated in line with recent number counts measurements from \citet{Geach2017} and \citet{Simpson2019}. The Cai-Negrello model over-predicts the S2LXS XMM-LSS observed number counts by a factor of approximately $2$, apart from in the brightest flux bins where the counts are consistent within the uncertainties.

The semi-analytic model {\sc shark} does not include a prescription for lensing or model the effects of source blending, and so, given these are known to be important influences, boosting the bright-end of the single-dish $850\,\mu{\rm m}$, we might expect this model to under-predict the number counts compared to the observed S2LXS XMM-LSS counts. \citet{lagos2019} combine the {\sc Shark} semi-analytical model \citep{lagos2018} with attentuation curves derived from {\sc Eagle} using the 3D dust radiative transfer code {\sc Skirt} \citep{trayford2017} to model the emission of galaxies from far-ultraviolet to far-infrared wavelengths. In Figure \ref{fig:number_counts} we show that the {\sc Shark} semi-analytical model over-predicts the bright-end of the number counts compared to the S2LXS XMM-LSS observed counts by up to a factor of $20$. 

The S2LXS XMM-LSS number counts are in broad agreement with the Cai-Negrello empirical model, the empirical model {\sc SIDES} (at flux densities below $30\,{\rm mJy}$) and the semi-analystic model {\sc galform}, which include gravitational lensing and/or the effects of source blending in a low-resolution single dish telescope beam, indicating that both may be important motivators for the observed over-abundance of bright single dish selected $850\,\mu{\rm m}$ sources.  With follow up ALMA observations (Project ID: 2022.1.01030.S) of the brightest sources in the S2LXS XMM-LSS catalogue we aim to place strong constraints on the contribution of instrinsically bright, lensed and blended sources to the bright-end of the $850\,\mu{\rm m}$ counts, vital to inform future models.    

\section{Summary}
  
We have presented the $850\,\mu{\rm m}$ maps and catalogue for the James Clerk Maxwell Telescope SCUBA-2 Large Extragalactic Survey of the XMM-LSS field. The S2LXS XMM-LSS survey maps a contiguous area of $9\,{\rm deg}^{2}$ to a moderate depth of about $4\,{\rm mJy\,beam}^{-1}$. This is the largest extragalactic area at $850\,\mu{\rm m}$ mapped with the JCMT to date. The wide area of the S2LXS XMM-LSS survey allows us to detect ultra-bright ($S_{\rm 850\mu m}>15\,{\rm mJy}$), but intrinsically rare sources, with approximately three sources expected per square degree with deboosted flux densities ${>}15\,{\rm mJy}$. The S2LXS XMM-LSS ${>}5\sigma$ catalogue comprises $40$ sources, of which $11$ have deboosted fluxes above $15\,{\rm mJy}$. This is twice the number of bright ($S_{\rm 850\mu m}>15\,{\rm mJy}$) extragalactic sources detected in S2CLS.  We use the S2LXS XMM-LSS ${>}5\sigma$ catalogue to estimate the number counts at intrinsic flux densities above $7\,{\rm mJy}$, significantly reducing the Poisson errors on existing measurements \citep[e.g.,][]{Chen2013,Coppin2006,Geach2017,Simpson2019,shim2020}. We observe the distinctive upturn in the number counts that is expected to be caused by submillimetre emission from Galactic objects and gravitationally lensed high-redshift galaxies. We note that at least two sources in the S2LXS XMM-LSS $5\sigma$ catalogue are already known, lensed, high-redshift ($z\gtrsim3$) galaxies \citep{Bussmann2015,Geach2017}.  Follow up ALMA observations (Project ID: 2022.1.01030.S, PI: T. K. Garratt) of the brightest sources in the S2LXS XMM-LSS catalogue will be a key step in unravelling the nature of the sources that contribute to the over-abundance observed at the bright-end of the single dish $850\,\mu{\rm m}$ number counts.  With the exquisite resolution of ALMA we aim to constrain the relative contributions of source blending, lensing and intrinsically bright sources to the bright-end of the $850\,\mu{\rm m}$ number counts; measure the redshift distribution for these rare, highly luminous galaxies; and explore whether a simple flux cut at $850\,\mu{\rm m}$ is an effective selection method to identify high-redshift, gravitationally lensed galaxies (Garratt et al. in prep).
                 
\section*{Acknowledgements}

We thank Matthieu B\'ethermin, Claudia del P. Lagos, Zhen-Yi Cai and Mattia Negrello for providing their number counts.  The James Clerk Maxwell Telescope is operated by the East Asian Observatory on behalf of The National Astronomical Observatory of Japan; Academia Sinica Institute of Astronomy and Astrophysics; the Korea Astronomy and Space Science Institute; the National Astronomical Research Institute of Thailand; Center for Astronomical Mega-Science (as well as the National Key R\&D Program of China with No. 2017YFA0402700). Additional funding support is provided by the Science and Technology Facilities Council of the United Kingdom and participating universities and organizations in the United Kingdom and Canada. Additional funds for the construction of SCUBA-2 were provided by the Canada Foundation for Innovation. The data presented in this paper was taken as part of Program ID M17BL001.  The authors wish to recognize and acknowledge the very significant cultural role and reverence that the summit of Maunakea has always had within the indigenous Hawaiian community.  We are most fortunate to have the opportunity to conduct observations from this mountain. 

TKG acknowledges support from a UK Science and Technology Facilities Council studentship. JEG is supported by a Royal Society University Research Fellowship. KEKC is supported by a Royal Society Leverhulme Senior Research Fellowship (SRF/R1/191013). KEKC and MF acknowledge support from STFC (grant number ST/R000905/1). C.-C.C. acknowledges support from the National Science and Technology Council of Taiwan (NSTC 109-2112-M-001-016-MY3 and 111-2112-M-001-045-MY3), as well as Academia Sinica through the Career Development Award (AS-CDA-112-M02). HD acknowledges financial support from the Agencia Estatal de Investigaci\'on del Ministerio de Ciencia e Innovaci\'on (AEI-MCINN) under grant (La evoluci\'on de los c\'iumulos de galaxias desde el amanecer hasta el mediod\'ia c\'osmico) with reference (PID2019-105776GB-I00/DOI:10.13039/501100011033) and acknowledges support from the ACIISI, Consejer\'ia de Econom\'ia, Conocimiento y Empleo del Gobierno de Canarias and the European Regional Development Fund (ERDF) under grant with reference PROID2020010107. This work has been partially supported by the Japan Society for the Promotion of Science (JSPS) Grants-in-Aid for Scientific Research (19H05076 and 21H01128).  This work has also been supported in part by the Sumitomo Foundation Fiscal 2018 Grant for Basic Science Research Projects (180923), and the Collaboration Funding of the Institute of Statistical Mathematics ``New Development of the Studies on Galaxy Evolution with a Method of Data Science".  
MJM acknowledges the support of the National Science Centre, Poland through the SONATA BIS grant 2018/30/E/ST9/00208.  HSH acknowledges the support by the National Research Foundation of Korea (NRF) grant, No. 2021R1A2C1094577, funded by the Korea government (MSIT). HS acknowledges the support from the National Research Foundation of Korea (NRF) grant, Nos. 2021R1A2C4002725 and 2022R1A4A3031306, funded by the Korea government (MSIT). YQX acknowledges support from NSFC grants (12025303 and 11890693). 

\section*{Data Availability}
We release the S2LXS XMM-LSS ${>}5\sigma$ catalogue as part of this publication, along with the $850\,\mu{\rm m}$ maps for use by the community. The data are available at the DOI https://doi.org/10.5281/zenodo.7371860





\bibliographystyle{mnras}
\bibliography{REF} 





\bsp	
\label{lastpage}
\end{document}